\documentclass[sn-mathphys,Numbered]{sn-jnl}
\usepackage{graphicx}
\usepackage{paralist}
\usepackage{multirow,multicol}
\usepackage{amsmath,amssymb,amsfonts,mathdots}
\usepackage{mathrsfs}
\usepackage{xcolor}

\newcommand{\cor}[1]{\textcolor{black}{#1}}
\newcommand{\cb}[1]{\textcolor{black}{#1}}
\newcommand{\minorrev}[1]{\textcolor{black}{#1}}
\usepackage{hyperref}
\usepackage{colortbl}
\usepackage{textcomp}
\usepackage{manyfoot}
\usepackage{footnote}
\usepackage{algorithm}
\usepackage{algorithmicx}
\usepackage{algpseudocode}
\usepackage{listings}
\usepackage{makecell}
\usepackage{xspace}
\usepackage{float}
\usepackage{longtable}
\usepackage{subcaption}
\floatstyle{plaintop}
\restylefloat{table}
\usepackage{multirow}
\usepackage{url}
\usepackage{mathtools}
\usepackage[utf8]{inputenc}
\usepackage{booktabs}
\usepackage{longtable}
\usepackage{tablefootnote}
\usepackage{nicematrix, tikz}
\usepackage{adjustbox}
\usepackage{anyfontsize}
\usepackage{bookmark}
\usepackage{natbib} 
\usepackage{comment}
\usepackage{mdframed}
\usepackage{mdframed}
\usepackage{enumitem}
\newmdenv[
  linewidth=1.5pt,
  linecolor=black,
  topline=false,
  bottomline=false,
  rightline=false,
  leftmargin=0pt,
  innertopmargin=0pt,
  innerbottommargin=0pt,
  innerleftmargin=5pt,
  innerrightmargin=0pt
]{myquote}
\usepackage[normalem]{ulem}
\usepackage[most]{tcolorbox}
\newtcolorbox{summarybox}[1][]{
  colback=gray!10,
  colframe=gray!135,
  arc=1mm,
  boxrule=0.5pt,
  left=5pt,
  right=5pt,
  top=5pt,
  bottom=5pt,
  title=#1
}
\newtcolorbox{takeawaybox}{
  colback=gray!10,
  colframe=gray!40,
  boxrule=0.3pt,
  arc=1mm,
  left=4pt, right=4pt,
  top=2pt, bottom=2pt,
}
\newtcolorbox{definitionbox}[1][]{
  colback=black!1,
  colframe=black!75,
  boxrule=0.5pt,
  arc=2mm,
  left=6pt, right=6pt,
  top=4pt, bottom=4pt,
  title={#1},
  fonttitle=\bfseries
}
\let\cline\cmidrule

\newcommand{\quoteFont}[1]{\emph{\textcolor{black}{#1}}}  % Italicized, bold, and blue color for the quotes

\newcommand{\total}{22\xspace}
\newcommand{\questions}{27\xspace}
\usepackage{nicematrix, tikz}
\usetikzlibrary{positioning}

\begin{document}

\title{\hspace{2cm}Security Debt in Practice:  \\ Nuanced Insights from Practitioners}

\author*[1]{\fnm{Chaima} \sur{Boufaied}}\email{chaima.boufaied@ucalgary.ca}

\author[2]{\fnm{Taher} \sur{A. Ghaleb}}\email{taherghaleb@trentu.ca}

\author[3]{\fnm{Zainab} \sur{Masood}}\email{zmasood@psu.edu.sa}

\affil*[1]{\orgname{University of Calgary}, \country{Canada}}
\affil[2]{\orgname{Trent University}, \country{Canada}}
\affil[3]{\orgname{Prince Sultan University}, \country{Saudi Arabia}}

\abstract{With the increasing reliance on software and automation nowadays, tight deadlines, limited resources, and prioritization of functionality over security can lead to insecure coding practices.
When not handled properly, these constraints cause unaddressed security vulnerabilities to accumulate over time, forming Security Debts (\textit{SDs}).
Despite their critical importance, there is limited empirical evidence on how software practitioners perceive, manage, and communicate SDs in real-world settings.
In this paper, we present a qualitative empirical study based on semi-structured interviews with \total software practitioners across various roles, organizations, and countries. We address four research questions: i) we assess software practitioners' knowledge of SDs and awareness of associated security risks, ii) we investigate their behavior towards SDs, iii) we explore common tools and strategies used to mitigate SDs, and iv) we analyze how security risks are communicated within teams and to decision makers.
We observe variations in how practitioners perceive and manage SDs, with some prioritizing delivery speed over security, while others consistently maintain security as a priority. Our findings emphasize the need for stronger integration of security practices across the Software Development Life Cycle (SDLC), more consistent use of mitigation strategies, better balancing of deadlines, resources, and security-related tasks, with attention to the \textit{Confidentiality, Integrity, and Availability} (CIA) triad.
}

\keywords{Security debt, unaddressed security vulnerability, qualitative study, empirical study, software security}

\maketitle

\section{Introduction}\label{sec:introduction}
In modern software development,
meeting tight delivery deadlines and fulfilling customer needs by supporting all functional requirements are among the top priorities for organizations~\cite{kruke2024defining}, regardless of their technical resources (e.g., lack of security tools~\cite{edmundson2022sans,ali2025assessing}) or organizational constraints (e.g., limited expertise, tight deadlines, and budget limitations)~\cite{coetzer2024managing,tondel2022influencing,rindell2019managing,ali2025assessing,kruke2024defining}. 

This industry-wide prioritization, combined with resource limitations and skill gaps, often comes at the expense of essential security measures, including security metrics and standards, which are frequently overlooked or deprioritized~\cite{zhao2024identifying}. As a result, unaddressed security vulnerabilities tend to accumulate over time, forming what is known as Security Debts (\textit{SDs}\cb{)}, also referred to as \textit{unresolved or unaddressed security vulnerabilities} throughout the paper, and sometimes called cyber SDs~\cite{coetzer2024managing}. % or unpatched vulnerabilities~\cite{ali2025assessing})}.}
SD is a specific subset of technical debt~\cite{rindell2019managing} that arises from postponed security-related tasks during software development. These unresolved vulnerabilities can lead to security breaches, compromise reliability (e.g., attacks such as unauthorized access leading to impersonation and bypassed authentication), and degrade maintainability (e.g., due to increased complexity resulting from unpatched security flaws and increased effort required for updates), ultimately risking system failure~\cite{ali2025assessing,rindell2019managing,coetzer2024managing}.
To mitigate such risks, it is widely recommended~\cite{rindell2019managing, zhao2024identifying, maier2017towards, zarour2020software,tondel2022influencing,ali2025assessing,kudriavtseva2024you,odera2023security,siavvas2022technical,theurich2023practices} to integrate security early across the Software Development Life Cycle (SDLC), including requirement elicitation, specification, architecture, implementation, and testing phases. This is encouraged across different software development methodologies and practices such as DevSecOps~\cite{rindell2019managing,zhao2024identifying} and Agile-based methods~\cite{maier2017towards,tondel2022influencing,theurich2023practices}.

Proactive SD prevention involves structured security analysis methods (e.g., threat modeling, risk assessment), supported by secure coding practices, code reviews, and early use of analysis and dependency management tools~\cite{konev2022survey,edmundson2022sans,maier2017towards,kudriavtseva2024you,grossman2007xss,edbert2023exploring}.
\cb{Security-specific static analyzers (e.g., Bandit~\cite{bandit}, Semgrep~\cite{semgrep}, Checkmarx~\cite{checkmarkx}) that detect vulnerability patterns directly in source code, and security testing techniques such as penetration testing and dynamic analysis, further support SD detection before release. }
Indicators such as code smells and unresolved bugs also serve as useful proxies in CI/CD and DevOps pipelines~\cite{siavvas2022technical}, and have been used to train machine learning models to predict security risks. Recent tools, including automation and conversational assistants, aim to support security throughout development~\cite{voggenreiter2024automated,tony2022conversational}. Despite these efforts, challenges continue to exist in generating meaningful metrics, aligning support from developers, and achieving early integration across SDLC stages~\cite{kudriavtseva2024you,zhao2024identifying}.

Despite the critical importance of SD and the research efforts dedicated to it, \cb{existing empirical work on how software practitioners perceive, manage, and communicate SD in real-world settings remains limited in scope and context, often focusing on a single corporation~\cite{kruke2024defining} or specific type(s) of technical debt (self-admitted, intentional) in open-source settings~\cite{diaz2024can}.}
To address this gap, we present a qualitative empirical study aimed at developing a deep, practice-informed understanding of SD from the perspective of software practitioners. We conducted semi-structured interviews with \total practitioners in various roles (e.g., software developer, software engineer, and data scientist) from organizations across ten countries. Our study investigates: i) how practitioners conceptualize and make sense of SD in their assigned projects, ii) how they behave toward SD during development, iii) what tools and strategies they use to mitigate SD under delivery constraints, and iv) how SD-related risks are communicated within and across teams. \cb{Our study extends beyond existing work~\cite{kruke2024defining,diaz2024can} by: (i) recruiting practitioners from diverse organizations, roles, domains, and countries; (ii) examining four complementary dimensions of SD (understanding SDs and their causes, behavioral analysis, mitigation strategies, and communication practices); (iii) covering both intentional and unintentional SDs; and (iv) systematically investigating practitioner behavior, tool usage including AI adoption, and risk communication.}
Our study offers practitioner-grounded insights into the realities of managing SDs in industrial settings, revealing how trade-offs, organizational practices, and tool limitations shape the way security risks are perceived, addressed, and communicated.

\vspace{4pt}
\noindent\textbf{Paper Contributions.}
This paper makes the following contributions:
\begin{itemize}
    \item[-] We conduct a qualitative study with 22 practitioners across ten countries to understand how SDs are perceived and handled in practice.
    \item[-] \cb{We propose a practitioner-informed definition of SD capturing design/implementation flaws, gaps between current and ideal security, and postponed security decisions.}
    \item[-] We empirically derived themes %empirically grounded themes 
    that capture how practitioners manage SDs under deadline pressure, mitigate them using technical and organizational strategies, and communicate associated risks.
    \item[-] We highlight actionable insights and implications to support more consistent, informed, and secure development practices.
\end{itemize}

\vspace{4pt}
\cb{\noindent\textbf{Summary of Key Findings.}
Our study reveals that practitioners define SDs either as system flaws prone to exploitation or as deferred security issues, with time pressure and weak security requirements as main causes. Senior and security-focused participants assess severity, prioritize by risk, and consider both short- and long-term impacts, while junior or non-security participants treat all SDs equally and rely on overtime or direct remediation, sharing the broader consensus that SDs are primarily long-term risks though this pattern is most evident in RQ1 and RQ2 and does not extend to mitigation (RQ3) or tracking practices (RQ4). Mitigation practices are inconsistent, formal processes and tool awareness are limited, and AI adoption is cautious. Communication about SD risks is informal and varies widely.
These findings have implications for research, practice, and education. For research, they challenge assumptions (e.g., tool unavailability as the main cause) and highlight overlooked factors like intentional trade-offs and informal communication. For practitioners, they stress embedding security discussions, standardizing mitigation practices, and fostering shared responsibility, while for education, they underscore teaching SD concepts and impacts, using hands-on tools, case studies, and team projects to prepare students to manage SDs effectively from the start of their careers.}

\vspace{4pt}
\noindent\textbf{Paper Structure.}
The rest of this paper is structured as follows: Section~\ref{sec:background} introduces relevant background concepts. Section~\ref{state} reviews related work. Section~\ref{sec:methodology} outlines the research methodology used for our qualitative analysis. Section~\ref{sec:findings} presents the findings of our research questions, based on the responses of interview participants. In Section~\ref{sec:discussion}, we position our study within the context of prior research, discuss the practical implications of our findings, analyze threats to validity, highlight limitations of our approach, and suggest directions for future work. Finally, Section~\ref{sec:conclusion} concludes the paper.

\section{Background}
\label{sec:background}

This section provides an overview of SDs by defining the concept, outlining its types, explaining how SDs accumulate across different stages of the Software Development Life Cycle (SDLC), and highlighting associated risks and mitigation techniques.

% DEF
\subsection{Security Debts: Definition and Types}

Software security, typically addressed through well-defined security requirements, involves implementing measures such as sensitive data protection, access control, and encryption~\cite{mcgraw2004software}. In many fast-paced software development environments, however, delivering functional requirements requested by clients often takes priority over enforcing security practices~\cite{zarour2020software}. As a result, code is frequently rushed into production without adequate security reviews~\cite{rindell2019managing}, especially under tight deadlines~\cite{coetzer2024managing, diaz2024can, kruke2024defining,maier2017towards,odera2023security,rindell2019managing,theurich2023practices,tondel2022influencing}.

For example, a developer implementing a login feature may overlook critical security measures such as password encryption or input validation. These omissions can leave the software vulnerable to code-level attacks like \textit{Cross Site Scripting} (XSS)~\cite{grossman2007xss}. Such weaknesses, known as security vulnerabilities, are flaws in 
design or implementation that attackers can exploit to compromise 
a system's Confidentiality, Integrity, or Availability (the CIA 
triad~\cite{whitman2004principles}). \cb{\textit{Confidentiality} ensures 
that sensitive data is accessible only to authorized parties; 
\textit{Integrity} ensures that data and system behavior remain unaltered 
by unauthorized actions; and \textit{Availability} ensures that systems 
and data remain accessible when needed. Security vulnerabilities 
can compromise one or more of these properties, for example, 
SQL injection may expose sensitive data (confidentiality), enable 
unauthorized data modification (integrity), or disrupt service 
through resource exhaustion (availability).}

When these vulnerabilities are left unresolved due to reasons such as deadlines, limited expertise, budget constraints~\cite{rindell2019managing,coetzer2024managing,tondel2022influencing,ali2025assessing,kruke2024defining}, poor requirements, weak design, or insufficient testing~\cite{zarour2020software}, they begin to accumulate and form SDs. According to Veracode's State of Software Security 2024 report~\cite{veracode2024}, SDs affect 42\% of software applications and 71\% of organizations\cb{, with 45.9\% of organizations carrying critical vulnerabilities. These figures underscore the widespread and severe nature of SD accumulation in practice, reinforcing the need for empirical research into how practitioners perceive and manage such vulnerabilities.}

As a type of technical debt~\cite{rindell2019managing}, SDs result from limited or postponed security practices. They may be i) \emph{intentional}, reflecting trade-offs for faster delivery, as seen in TODO comments or pending issues~\cite{diaz2024can}) or ii) \emph{unintentional}, stemming from lack of awareness or training~\cite{rindell2019managing,diaz2024can}). \\
Overall, SDs can manifest across several layers of a software system: the network, user behavior, and application layers~\cite{ali2025assessing}. At the network layer, for example, missing protections like SYN cookies may enable SYN flooding attacks~\cite{bellovin1989security}. At the user level, weak access control can expose session cookies, allowing impersonation~\cite{woschek2015owasp}. At the application level, SDs often result from insecure programming practices such as weak passwords, lack of input validation, or missing bounds checks, leading to vulnerabilities like SQL injection~\cite{forristal1998sql}, Cross-Site Scripting (XSS)~\cite{microsoft2000xss}, or buffer overflows~\cite{bufferOverflow}. These unresolved flaws accumulate over time, increasing system risk. In this study, we focus on SDs at the application level; network and user-level SDs are out of scope.

\subsection{Security Debts across the SDLC: Risks and Mitigation}

SDs may emerge at various stages of the SDLC, especially in agile and DevOps-based workflows~\cite{daneva2018security, edmundson2022sans, odera2023security, rindell2019managing, theurich2023practices, tondel2022influencing, zhao2024identifying, maier2017towards, voggenreiter2024automated}. An empirical study analyzing Common Weakness Enumerations (CWEs) found that 16.9\% of weaknesses stem from requirement-related issues, while 72.1\% are linked to design and implementation flaws~\cite{diaz2024can}, which underscores the importance of early security integration in SDLC~\cite{humayun2022security}.

In the \emph{design} phase, skipping secure design principles, such as least privilege or separation of concerns~\cite{saltzer1975protection}, may result in vulnerable architectures. In the \emph{implementation} phase, using deprecated libraries or poorly documented code increases exposure to security risks~\cite{coetzer2024managing}. Integrating secure practices early can help detect, prioritize, and mitigate SDs~\cite{zhao2024identifying, ali2025assessing, martinez2021security}, reducing the risk of future incidents~\cite{coetzer2024managing}. Awareness of typical attacks (e.g., phishing, SQL injection [SQLi], and cross-site scripting [XSS]) supports proactive vulnerability prevention~\cite{ali2025assessing}.

In the \emph{requirements} phase, identifying critical assets informs priority-setting~\cite{ali2025assessing, daneva2018security}. Effective security requirements often rely on threat modeling~\cite{konev2022survey, scandariato2015descriptive}, risk analysis methods like CORAS~\cite{lund2011guided}, or models such as DREAD~\cite{kierkegaard1957concept}. These techniques assist in early threat identification~\cite{ali2025assessing, coetzer2024managing, edmundson2022sans, kudriavtseva2024you}.

In the \emph{implementation} phase, secure coding practices (e.g., input validation, password management, defense in depth) are essential~\cite{graff2003secure, securecoding_owasp}. Regular code and configuration reviews serve as effective strategies for catching and addressing vulnerabilities before release~\cite{ali2025assessing, huopio2020quest}.

In sum, embedding security into each phase of the SDLC, especially early in requirements, design, and implementation phases, is vital to preventing SD accumulation and building resilient, trustworthy software~\cite{daneva2018security, kudriavtseva2024you, zhao2024identifying, zarour2020software}.

\section{Related Work}
\label{state}

\cb{To investigate how SDs are handled in practice, we conducted a targeted literature search using a query \footnote{Query: ((“security debt” OR “unaddressed security vulnerability”) AND (“SDLC” OR “software development life cycle”) AND (“detection” OR “mitigation” OR “prioritization” OR “risk”) AND “practitioners”).}. This search identified relevant studies on unaddressed security issues across the SDLC, including how practitioners manage SDs and assess associated risks. Prior work can be broadly grouped into three areas: (1) causes of SDs in practice, (2) challenges in managing SDs across the SDLC, and (3) tools and techniques supporting their identification and prioritization. Below, we review these aspects and highlight key gaps addressed by our study.}

\cb{\subsection{Causes of SDs  In Practice}}
\cb{Prior research identifies lack of security awareness as a primary cause of SDs among software practitioners~\cite{kruke2024defining, ali2025assessing, zhao2024identifying}. 
These studies highlight challenges such as difficulty distinguishing between vulnerabilities and SDs, insufficient training, and limited understanding of secure development practices. For example, one study~\cite{kruke2024defining} provides a comprehensive analysis of these issues, emphasizing how gaps in developer knowledge contribute to the accumulation of unresolved security problems.}

\cb{In addition, organizational constraints such as time pressure, tight deadlines, and poor cross-team communication further contribute to SDs~\cite{coetzer2024managing, rindell2019managing, odera2023security}.
Other studies point to technical and process-related issues, including inadequate security tools, poorly defined requirements, and reliance on legacy or third-party components~\cite{edmundson2022sans,rindell2019managing, maier2017towards}.
One study \cite{tondel2022influencing} offers an in-depth case study of SD management within a single organization, highlighting how resource limitations and local practices influence how security issues are handled, although its findings may be limited in generalizability.}

\cb{Overall, the literature indicates that SDs arise from a combination of human, organizational, and technical factors. However, many studies focus on these aspects in isolation, providing limited insight into how they interact in practice and how SDs are managed across different contexts.}

\cb{\subsection{Challenges in Managing SDs}}
\cb{Research highlights several practices for managing and mitigating security across the SDLC. Early-stage approaches emphasize integrating security from the outset, including secure requirements engineering and developer training, to reduce the likelihood of introducing vulnerabilities~\cite{ali2025assessing, zarour2020software}. In the design phase, techniques such as threat modeling are used to proactively identify and address potential risks before implementation~\cite{konev2022survey}. During development, practices such as secure coding and code reviews are widely adopted to detect and prevent vulnerabilities before deployment~\cite{edmundson2022sans,rindell2019managing}. In addition, teams often rely on coordination mechanisms, including cross-team collaboration and DevOps practices, to manage security issues more effectively in fast-paced development environments.}

\cb{Prior studies report multiple challenges in consistently applying security practices across SDLC phases. These include involving developers instead of analysts in drawing misuse cases~\cite{alexander2003misuse}, research emphasizes the importance of integrating security from the earliest stages, particularly requirement elicitation~\cite{zarour2020software, ali2025assessing, daneva2018security}. This is especially difficult in Agile development, where requirements evolve continuously~\cite{theurich2023practices,edmundson2022sans,maier2017towards,odera2023security,rindell2019managing,tondel2022influencing}. Recommended approaches include investing in developer security education and using artifacts like abuse case diagrams~\cite{daneva2018security,mcgraw2004software}.
In the design phase, applying threat modeling~\cite{konev2022survey,scandariato2015descriptive,maier2017towards}, dependency checking~\cite{maier2017towards}, and structured risk analysis techniques such as CORAS~\cite{lund2011guided} and DREAD~\cite{kierkegaard1957concept} helps proactively manage risks. In development, secure coding principles~\cite{graff2003secure,securecoding_owasp} and code reviews are widely endorsed~\cite{edmundson2022sans,rindell2019managing,maier2017towards,kudriavtseva2024you,diaz2024can,huopio2020quest} to prevent vulnerabilities before code reaches production.}

\cb{\subsection{Tools and Techniques for SD Management}}
\cb{Prior research has also proposed various tools and techniques to support the identification and mitigation of SDs. Some studies use technical debt indicators (e.g., code smells, duplication, unresolved bugs) as proxies for assessing SDs, although these approaches provide limited security-specific signals~\cite{rindell2019managing,siavvas2022technical}.}

\cb{Static analysis and security testing techniques offer more direct support for vulnerability detection. Security-oriented linters and Static Application Security Testing (SAST) tools (e.g., Bandit~\cite{bandit}, Semgrep~\cite{semgrep}, and Checkmarx~\cite{checkmarkx}) scan source code for known vulnerability patterns, enabling early detection. Dynamic approaches, including penetration testing~\cite{pentesting}, fuzz testing~\cite{liang2018fuzz}, and Dynamic Application Security Testing (DAST)~\cite{pan2019interactive} with tools such as OWASP ZAP~\cite{owasp_zap}, complement static analysis by identifying runtime vulnerabilities that may not be visible in code ~\cite{zhao2024identifying}.}

\cb{Despite their complementary strengths, studies show that individual SAST tools detect between 11\% and 26\% of known vulnerabilities~\cite{semgrep}, and adoption remains inconsistent across development teams, highlighting a persistent gap between available tooling and real-world SD mitigation practice. More advanced approaches, such as machine learning-based prediction models [12] and DevSecOps tools (e.g., Security Flama~\cite{voggenreiter2024automated} and chatbot-based assistants~\cite{tony2022conversational,SKFChatbot}), aim to support prioritization and remediation. However, these tools often provide fragmented support and may not align well with real-world development workflows. For example, Conversational DevBots like SKF chatbot~\cite{tony2022conversational,SKFChatbot} have been evaluated for their ability to assist developers in resolving vulnerabilities during development. While some participants in those studies found chatbots helpful, others preferred manual searches due to generic or repetitive responses. The Challenge-Practice-Tool-Metric (CPTM) model~\cite{zhao2024identifying} provides a DevSecOps-aligned framework, mapping challenges to tools and practices across the SDLC. However, recent research shows that security metrics remain underexplored~\cite{kudriavtseva2024you}, often misaligned with industry practice and lacking interpretability and guidance for early-phase decisions.}

\cb{Overall, while existing tools provide valuable support, their practical adoption and effectiveness remain inconsistent, highlighting a gap between available tooling and real-world SD mitigation practices.} \\

\subsection{Relevant Empirical Studies on SDs}
Overall, the literature shows that SDs are not solely technical in nature; they stem from organizational, process, and knowledge limitations. Despite the availability of tools and practices, consistent challenges persist in metrics, developer support, early integration, and security prioritization across the SDLC.
Only a few empirical studies have explored SDs in practice. 
 
\begin{table}[ht]
    \centering
    \caption{Summary of Relevant Empirical Studies on SDs}
    \label{tab:related}
    \resizebox{1\linewidth}{!}{
        \begin{tabular}{p{2.9cm} p{2cm} p{5.1cm} p{3.5cm}}
        \hline
        \textbf{Study} & \textbf{Empirical Method} & \textbf{Contribution(s)} & \textbf{Stakeholders Involved} \\
        \hline
        \citet{diaz2024can} & Online ~~~~~survey & Analyzed security risks linked to self-admitted technical debt (i.e., intentionally introduced flaws) in open-source projects. & 222 open-source developers \\
        \hline
        \citet{kudriavtseva2024you} & Practitioner workshop & Compared software security metrics from academia and industry to evaluate SDLC coverage and alignment. & 7 participants: developers, security specialists, project managers \\
        \hline
        \citet{tondel2022influencing} & Interviews & Explored how security experts influence prioritization decisions in agile development environments. & 1 organization: developers, product owners, managers, security officer \\
        \hline
        \citet{kruke2024defining} & Semi-structured interviews & Investigated practitioners’ understanding of SDs, their relation to vulnerabilities and technical debt, and how SDs accumulate. & 26 practitioners from a global software corporation \\
        \hline
        \citet{tony2022conversational} & Interviews and surveys & Evaluated the usefulness of the SKF chatbot (a DevBot) in helping students fix security vulnerabilities during development. & 15 master's students with basic to intermediate security knowledge \\
        \hline
        \citet{voggenreiter2024automated} & Unstructured interviews & Assessed the effectiveness of Security Flama in reducing security issues in an industrial DevOps setting, highlighting the importance of communication strategies. & 5 developers and stakeholders in industrial DevOps projects \\
        \hline
        \textbf{Our Study} & Semi-structured interviews & Broad investigation into how practitioners understand, manage, and communicate SDs, including behaviors, tools, and mitigation strategies. & \total practitioners from various industries and countries \\
        \hline
        \end{tabular}
    }
\end{table}

Table~\ref{tab:related} summarizes these works, listing their \textit{Empirical Method}, key \textit{Contribution(s)}, and \textit{Stakeholders}. \cb{Though some studies share methodological similarities with ours, such as the use of semi-structured interviews in~\cite{kruke2024defining} and~\cite{tondel2022influencing}, or the focus on practitioner perspectives in~\cite{kudriavtseva2024you} and~\cite{diaz2024can}, they differ from our work in important ways. \citet{kruke2024defining} conducted interviews within a single global software corporation, limiting the diversity of perspectives captured, and focused primarily on understanding SDs and their accumulation without examining mitigation tools, AI adoption, or communication practices. \citet{diaz2024can} surveyed open-source developers about self-admitted technical debt, restricting their scope to intentional debt in open-source contexts.}
Other studies are either limited in scale~\cite{kudriavtseva2024you,tondel2022influencing,tony2022conversational,voggenreiter2024automated} or narrowly focused on specific tools such as the SKF chatbot~\cite{tony2022conversational} and Security Flama~\cite{voggenreiter2024automated}, or on comparing security metrics for SDLC coverage~\cite{kudriavtseva2024you}. \cb{In contrast, our study offers three distinct advances over this body of work: (i) breadth across four research questions covering practitioner understanding, behavior, mitigation strategies, and communication of SD risks; (ii) diversity across \total practitioners spanning multiple roles, domains, experience levels, and ten countries; and (iii) coverage of both intentional and unintentional SDs in industrial settings. To our knowledge, this is the first empirical study to simultaneously examine all four of these dimensions of SD in practice.}

\section{Methodology}\label{sec:methodology}

We conducted a qualitative empirical study involving \total software practitioners from different organizations, roles, \cb{and expertise levels,} across ten countries to explore their perspectives on SD. 
\begin{figure}[H]
\centering
\resizebox{.65\linewidth}{!}{
\begin{tikzpicture}[
    rect/.style={draw, rectangle, minimum width=3cm, minimum height=1cm},
    dashedgroup/.style={draw, rectangle, dashed, minimum width=1cm, minimum height=1m, inner sep=10pt},
    annot/.style={rotate=90, anchor=east},
    interview/.style={fill=gray!20}
]
% Main rectangles
\node[rect] (A0) at (0,2) {Identify Problem \& Objectives}; % New top node
\node[rect, below=0.7cm of A0] (A) {Define Research Questions};

% Arrows connecting nodes
\draw[->] (A0.south) -- (A.north); % Connect new node to A

% Interview process (C1, C2, C3)
\node[rect, below=1cm of A, interview] (C1) {  
\hspace{.35cm}\begin{minipage}{8cm}
        \begin{itemize}
            \setlength\itemindent{.3cm} % Adjust the indent
            \item \small{Design interview guide (semi-structured)}
            \item \small{Recruitment of \total participants}
            \item \small{Prepare logistics (call scheduling and setup)}
            \item \small{Obtain \cb{informed} consent
            %(ethical approval)
            }
        \end{itemize}
    \end{minipage}};
    
\node[rect, below=1cm of C1, interview] (C2) {  
\hspace{.35cm}\begin{minipage}{8cm}
        \begin{itemize}
            \setlength\itemindent{.3cm} % Adjust the indent
            \item \small{Conduct pilot interviews}
            \item \small{Conduct individual interviews}
        \end{itemize}
    \end{minipage}
};
    
\node[rect, below=1cm of C2, interview] (C3) {  
\hspace{.35cm}\begin{minipage}{8cm}
        \begin{itemize}
            \setlength\itemindent{.3cm} % Adjust the indent
            \item \small{Collect responses}
            \item \small{Analyze data (Thematic Analysis)}
            \item \small{Report findings and actionable insights}
        \end{itemize}
    \end{minipage}
}; 

% Connecting arrows for interview steps
\draw[->] (C1.south) -- (C2.north);
\draw[->] (C2.south) -- (C3.north);

% Dashed boundary box
\draw[dashed] (-4.7,-.4) rectangle (4.7,-7.7);
\draw[->] (A.south) -- (C1.north) node[midway, fill=white] {Pre-Interview};
\draw[->] (C1.south) -- (C2.north) node[midway, fill=white] {Interview};
\draw[->] (C2.south) -- (C3.north) node[midway, fill=white] {Post-Interview};

\end{tikzpicture}
}
\vspace{5pt}
\caption{Overview of our Research Methodology
% (Pre-Interview, Interview, and Post-Interview Phases)
}
\label{process}
\end{figure}

\figurename~\ref{process} gives an overview of our research methodology, which includes the \textit{pre-interview phase} for the selection and preparation of participants, the \textit{interview phase} for data collection, and the \textit{post-interview phase} for transcription collection, qualitative data analysis, and validation.

\subsection{Research Questions}\label{objAndRQs}
Understanding the awareness of software practitioners of the criticality and impact of SD on software reliability and maintainability is as important as investigating why SD emerges and accumulates over time. This is particularly important in high-quality software designed to meet industry standards and serve paying customers, even with significant investments in staff and resources within industries.
The goal of our study is fourfold: i) examine how practitioners conceptualize and make sense of SDs in projects assigned to them, ii) examine how software practitioners behave toward SDs in software development, iii) explore the techniques and tools they adopt to mitigate SDs under delivery constraints; and (iv) investigate how SDs are discussed and communicated across roles and teams.
To gain these insights, we address the following research questions:

\begin{itemize}
    \item \textbf{RQ1:} \textit{To what extent are software practitioners knowledgeable about SDs and their impact on software security?}
    
    \item \textbf{RQ2:} \textit{How do software practitioners behave towards SDs in software development?}
    
    \item \textbf{RQ3:} \textit{How do software practitioners mitigate SDs in software?}
    
    \item \textbf{RQ4:} \textit{How do software practitioners communicate the risks associated with SDs in software to teammates and decision makers?}
\end{itemize}

\subsection{Interview Guide Design}
We designed an interview guide comprising \questions questions, including 22 open-ended questions and five follow-up questions designed to encourage further elaboration. The guide starts with a few introductory questions to understand participants' work, roles and responsibilities, the product(s) they are involved with, and other demographic information. Then, it includes questions related to understanding, identifying, and managing SDs, along with their perceived impact.
A full list of questions can be found in the interview guide~\cite{guide}, provided as supplementary material.
Each of the aforementioned research questions $RQi$ (i = 1..4) is addressed through $x$ sub-questions ($Qj$, where $j = 1..x$), denoted as $RQi_{Qj}$. For instance, to answer RQ1, we asked the following five
%(x = 5)
open-ended sub-questions (also see the interview guide~\cite{guide}):

\begin{itemize}
    \item $RQ1_{Q1}$: How would you define an unaddressed security vulnerability in a software? \\
    Follow-up: Can you provide us with an example of an unaddressed security vulnerability?
    \item $RQ1_{Q2}$: What factors /common causes do you think contribute to unaddressed security vulnerabilities in software?
    \item $RQ1_{Q3}$: What potential code vulnerabilities do you think are most likely to arise due to accumulating unaddressed security vulnerabilities?
    \item $RQ1_{Q4}$: Do you consider weak security measures as short-term mistakes in the code or long-term risks?  Why? 
    \item $RQ1_{Q5}$: From your experience, what is the potential impact of any unaddressed security vulnerabilities (e.g., impact on software reliability and maintainability)?
\end{itemize}

The above sub-questions were crafted to translate RQ1 into actionable prompts. Each sub-question targets a key aspect of practitioners’ understanding and perception of SDs and associated risks. Specifically,
$RQ1_{Q1}$ asked participants to define unaddressed security vulnerabilities and provide examples (in the corresponding follow-up question), establishing their basic understanding of SDs.
$RQ1_{Q2}$ focused on the causes and factors behind such unaddressed security vulnerabilities, revealing practitioners’ insights into why SDs occur in software.
$RQ1_{Q3}$ explored the types of code vulnerabilities linked to unaddressed security issues, illustrating practitioners’ awareness of the risks these vulnerabilities pose to software at the code level.
$RQ1_{Q4}$ examined whether SDs are seen as a short-term mistake or a long-term risk, helping to understand how practitioners perceive the urgency and severity of SDs, and whether they view them as a temporary issue in the code or a lasting threat to software quality. Finally,
$RQ1_{Q5}$ addressed the perceived impact on software quality, exploring practitioners' views on how SDs affect key aspects such as reliability and maintainability.

Overall, these sub-questions directly reflect the core aspects of RQ1, enabling a clear understanding of practitioners’ definitions ($RQ1_{Q1}$), ii) views ($RQ1_{Q4}$), and awareness of the causes ($RQ1_{Q2}$), risks ($RQ1_{Q3}$), and impacts of SDs on software quality ($RQ1_{Q5}$).
We adopted the same strategy in aligning the remaining research questions (RQ2--RQ4; see Section~\ref{objAndRQs}) and the corresponding sub-questions, following the approach illustrated above for RQ1. We highlighted RQ1 as an illustrative example to avoid redundancy and improve readability while maintaining transparency in our methodological design. 

\subsection{Participants Recruitment}
 As shown in Table~\ref{tab:roles_experience}, our overview of \total participants (P1--P22 in Column `ID') captures a diverse range of roles (Column `Role') such as software developer, data scientist, software engineer, penetration tester and security program manager across a wide range of domains (Column `Domain'), including finance, healthcare, aerospace, social media, e-commerce, and emerging areas like smart contracts and federated learning through a combination of personal contacts and public outreach.
Specifically, we announced a call for interviews via a post on LinkedIn (similar to other studies such as~\cite{kudriavtseva2024you}), which helped us reach professionals beyond our immediate network. Participants are based in 10 different countries (Column `Country'), with career experience (Column `Exp (Years)') ranging from 1 to 15 years (average ~5.6 years), representing a balanced mix of four junior (1--2 years), eleven mid-level (3--5 years), four senior (6--10 years), and three highly senior (11--15 years) professionals, with boundary values assigned to the higher category. We use the four rankings to allow for nuanced comparison of how professional experience shapes participants' understanding and perspectives on SDs.

\cor{Out of the \total participants, eight held roles that are directly security-related, such as mid-level and highly Senior Security Consultant (P10 and P17, respectively), Penetration Tester (P12), and Senior Leadership roles (e.g., Security Manager P13, P14, and Associate Manager Security P16). While five additional participants are not primarily in security positions, they still engage with security requirements as part of their development responsibilities. For instance, roles such as Sr. Enterprise App Engineer (P9), Technical Product Manager (P20), and QA Engineer (P6) often integrate security considerations into broader engineering tasks. Similarly, software developers from different levels, such as mid-level (P19, P22), senior (P7), and highly senior like P18, although not in security roles, described handling authentication features, password encryption, and two-factor authentication implementation, demonstrating that they are actively involved in software security while working on functional requirements.}

\begin{table*}[!ht]
    \renewcommand{\arraystretch}{1.2}
    \centering
    \caption{Participants Demographics; ID=Participant ID; Country [LU=Luxembourg, NZ=New Zealand, PK=Pakistan, EG=Egypt, UK=United Kingdom, US=United States of America, CA=Canada, IR=Iran, YE=Yemen, TU= Tunisia]}
    % \vspace{-5pt}
    \label{tab:roles_experience}
    \setlength{\tabcolsep}{8pt}
    \adjustbox{max width=\textwidth}{
    \begin{tabular}{l l l c c c} 
        \hline
        \rowcolor{gray!50}
        \textbf{ID} & 
        \textbf{Role} & 
        \textbf{Domain} & 
        \textbf{Gender} & 
        \textbf{Exp (Years)} &
        \textbf{Country} \\
        \hline
        \rowcolor{gray!15}
        P1  & \raggedright Research Engineer & \raggedright Automotive Safety & M & 2--5 & LU \\
        
        P2  & \raggedright Research Scientist & \raggedright Satellite & M & 1--2 & LU \\
        \rowcolor{gray!15}
        P3  & \raggedright Software Engineer & \raggedright Advertising & M & 1--2 & IR \\
        
        P4  & \raggedright Software Engineer & \raggedright E-commerce & M & 2--5 & IR \\
        \rowcolor{gray!15}
        P5  & \raggedright Data Scientist & \raggedright Multiple  & M & 1--2 & PK \\
        
        P6  & \raggedright QA Engineer & \raggedright Multiple  & F & 1--2 & PK \\
        
        \rowcolor{gray!15}
        P7  & \raggedright Software Developer & \raggedright Financial & M & 5--10 & YE \\
        
        P8  & \raggedright Software Engineer & \raggedright Social media & M & 2--5 & US \\
        
        \rowcolor{gray!15}
        P9  & \raggedright Sr. Enterprise Application Engineer & \raggedright Aerospace & M & 2--5 & US \\
        
        P10 & \raggedright Sr. Security Consultant & \raggedright Financial, Medical, Infrastructure & M & 2--5 & NZ \\
        
        \rowcolor{gray!15}
        P11 & \raggedright Software Engineer & \raggedright Financial & M & 2--5 & CA \\
        
        P12 & \raggedright Penetration Tester & \raggedright Banking & M & 2--5 & PK \\
        \rowcolor{gray!15}
        P13 & \raggedright Security Program Manager & \raggedright Multiple domains & M & 2--5 & EG \\
        
        P14 & \raggedright Security Manager & \raggedright Financial & M & 5--10 & PK \\
        
        \rowcolor{gray!15}
        P15 & \raggedright Application Security Consultant & \raggedright Financial, Infrastructure & M & 5--10 & PK \\
        
        P16 & \raggedright Associate Manager Security & \raggedright Healthcare, Financial, Infrastructure & M & 10--15 & PK \\
        
        \rowcolor{gray!15}
        P17 & \raggedright Security Consultant & \raggedright Banking & M & 10--15 & UK \\
        
        P18 & \raggedright Senior Software Developer & \raggedright E-commerce & F & 10--15 & LU, TU \\
        
        \rowcolor{gray!15}
        P19 & \raggedright Software Developer & \raggedright Electronic Design Automation & F & 2--5 & CA \\
         
        P20 & \raggedright Technical Product Manager & \raggedright Multi-Family Industry & F & 5--10 & US \\
        
        \rowcolor{gray!15}
        P21 & \raggedright AI Consultant & \raggedright Smart contracts in fintech & F & 2--5 & US \\
         
        P22 & \raggedright Software developer & \raggedright Federated Learning in Healthcare & F & 2--5 & LU \\
        \bottomrule
    \end{tabular}
    }
\end{table*}

Our initial interviews (P1--P17 in the table) included only one female participant. To reduce this gender imbalance, we recruited five additional female participants (P18--P22). Such gender gaps are common in computer science, \cb{where most practitioners are men, despite the gradual increase in female participation~\cite{cheryan2017some}.}
\cb{More specifically,} a recent survey~\cite{diaz2024can} on self-admitted technical debt found only 13\% of participants were female. 

Our participant selection aligned with qualitative sampling principles, particularly purposive and referral-chain sampling strategies~\cite{baltes2022sampling} and 
enabled us to access information-rich participants with relevant domain experience, and ensured heterogeneity across roles, domains, and genders.
\cb{Finally, while our sample may not be statistically representative, it captures a broad range of perspectives essential for analytical generalization and theory-building in qualitative research. 
This approach was chosen to ensure diversity and relevance in our data, enabling richer insights into the phenomenon under study. However, as with most qualitative research, the findings may be influenced by the sample's composition and cannot be generalized.
} 

\subsection{Interview Procedure}\label{proc}
All \total interviews were conducted virtually via
Google Meet, Zoom, and Microsoft Teams. The interviews lasted between 30 and 40 minutes, \cb{with 27 questions (22 open-ended and 5 follow-ups) designed to fit this time frame. \cb{This duration was intentionally designed to balance participant engagement with sufficient depth of response, while also providing other researchers with a realistic time estimate for replicating our study.}
We actively managed time during interviews by encouraging participants to be concise, reminding them of the time when they took longer on certain questions, and offering to skip questions if they felt unable to answer. Follow-ups were also asked selectively, not for every single participant, depending on their answers to earlier questions. 
For instance, follow-up questions 3.2.1 and 3.2.2, which probed the effectiveness and trustworthiness of AI tools in mitigating unresolved security vulnerabilities, were only asked when participants confirmed using such tools in question 3.2.
}

\cb{We conducted six pilot interviews with six participants to refine the interview questions and improve clarity (see the interview process in \figurename~\ref{process}).
The interview questions were developed to address gaps in prior SD literature, which has predominantly lacked the practitioner perspective on how SD is defined, caused, and managed~\cite{kruke2024defining}, while organizational causes such as deadline pressure and unclear security responsibilities remain underexplored from practitioners' own accounts~\cite{odera2023security, tondel2022influencing, zarour2020software}, existing mitigation tools fail to support early SDLC stages or provide qualitative insights meaningful to practitioners~\cite{kudriavtseva2024you}, and little is known about how SD risks are effectively communicated to decision makers who deprioritized security in favor of functional delivery~\cite{diaz2024can, tondel2022influencing}.
Further, a key insight from the pilot interviews was that the term `security debt' was not widely recognized, even among software specialists such as testers and developers. Consequently, we rephrased all interview questions to use `unresolved security vulnerability' instead, ensuring semantic consistency and participant comprehension. Additionally, pilot participants spontaneously discussed their experiences with tool reliability and effectiveness, which was not explicitly covered in our initial interview guide. This led us to add Question 3.1.2: `How reliable are these tools/techniques at mitigating unresolved security vulnerabilities?' to systematically capture these insights. Finally, we added a reflection question at the end of the interview, inviting participants to share any concerns or additional experiences not covered by our questions, ensuring we captured unanticipated perspectives and experiences.
\cb{Overall,} these refinements enhanced question clarity and ensured participants could provide meaningful responses across different experience levels and roles.}

Though some interview questions were reformulated/conducted in French or Arabic to respect participants’ linguistic and cultural preferences, we ensured that key terms remained consistent and the semantics of each statement were accurately preserved to maintain consistency across the different interviews.
We did not conduct formal member-checking after the interviews, as participants were only involved once.
\cb{We did not conduct formal member checking after the interviews, as participants were only involved once and were not available for follow-up meetings. This was a deliberate acknowledgment of our participants' time constraints and their limited commitment beyond the initial interview session.}
However, our interaction with the participants was proactive. As we took notes, we consistently asked for clarification and encouraged participants to elaborate or correct us during the conversation to ensure that their views were accurately captured.

\subsection{Ethics and Consent}

Prior to the interviews, all participants provided informed consent during the pre-interview process (see \figurename~\ref{process}). We explicitly communicated that participation was voluntary and that all data collected, including participants’ identities and company names, would remain strictly anonymous and confidential. 
Participants were assured that their identities would not be referenced in any published materials of this research, and that data would be accessible only to the three authors of the paper. 
We also recorded audio only for seven participants who gave explicit consent. 
\cb{The remaining participants were not recorded. Their data was captured in the transcripts during each interview.}
This practice ensured adherence to ethical standards and therefore supported a transparent and comfortable interview setting.
To ensure accuracy and rigor, we regularly reviewed the data and recorded our analysis decisions to create a clear audit trail, while reflecting on our own perspectives.

\subsection{Data Saturation}\label{datasaturation}

We initially conducted 17 interviews and observed that after this point, no new themes were emerging. Participants’ responses began to repeat similar ideas, indicating that we had reached saturation. The point at which further interviews were unlikely to provide additional insights related to our main research questions about SDs. Adding five more female participants was mainly to improve gender representation in the sample and reduce the gender gap. However, additional interviews confirmed existing themes and did not introduce new findings. Therefore, we stopped conducting interviews after the 22nd participant. \cb{This sample size is consistent with established guidelines for qualitative research using thematic analysis, where 20--40 interviews are recommended for heterogeneous, multi-site studies~\cite{hagaman2017many,hennink2022sample}, and aligns with sample sizes in comparable software engineering interview studies~\cite{kruke2024defining,tondel2022influencing,baltes2022sampling}.}

\cb{Data saturation in this study was determined at the level of individual codes rather than at the broader theme level, in order to ensure that all distinct concepts in the collected transcripts were captured before grouping them into larger themes. This approach reflects the description of saturation in the literature, defined as the point where additional interviews fail to provide new information that would require adding new codes to the codebook or altering category properties~\cite{fusch2015we}. Therefore, any subsequent instances corresponding to existing codes were considered redundant and did not contribute new information.} 
\cb{Table~\ref{saturation} illustrates a sample of major dimensions we covered for each RQ, showing the participant for whom data saturation was reached (Column `Sat. P\#'), along with notes on subsequent contributions.}

\begin{table}[h!]
\centering
\caption{\cb{Example of code-level saturation points for RQ1--RQ4.}}
\label{saturation}
\begin{tabular}{m{.55cm}|m{4cm}|m{1cm}|m{6cm}}
\hline
\textbf{\cb{RQ}} & \textbf{\cb{Question Context}} & \textbf{\cb{Sat. P\#}} & \textbf{\cb{Notes on subsequent contributions}} \\
\hline
RQ1 & Root causes of SDs & P16 & Quotes from P17 to P22 exist in the codebook. \\
RQ2 & SD prioritization & P12 & Shared experience from P13--P22 fits existing codes.\\
RQ3 & SD mitigation practices & P21 & Last participant's quote led to a new  code.\\ 
RQ4 & SD identification during testing & P19 & P20--P22 did not share additional techniques/ processes. \\
\hline
\end{tabular}
\end{table}

\subsection{Data Analysis}\label{data_analysis}
As part of the post-interview steps (see \figurename~\ref{process}), we adopted a rigorous and \cb{transparent approach to data analysis by manually analyzing the collected interview data using} \textbf{Thematic Analysis (TA)~\cite{clarke2017thematic}}, which supports the systematic identification and organization of recurring patterns, referred to as \textit{themes}, across the responses collected from the \total interviews.
\cb{The first author conducted the majority of the analysis, including the initial familiarization with the data and the manual coding of the transcripts, and consulted with the second and third authors in cases of doubts or difficulties. More in detail, we followed an open coding process implemented through collaborative \textit{card sorting} sessions~\cite{spencer2009card}, during which the first author generated initial codes and subsequently met with the other co-authors to review, discuss, and reconcile interpretations.
}
Disagreements were resolved through discussion and consensus \cb{meetings}, and similar codes were refined and merged into shared themes.
We adhered to the six phases of TA outlined by~\citet{clarke2017thematic}
, given its flexibility and suitability for identifying patterns across qualitative data, while also enabling a nuanced interpretation grounded in participants’ experiences.
As depicted in Table~\ref{TAsteps}, each phase of the TA (Column `Step') involved specific inputs, activities, and outputs, allowing for a structured, transparent, and rigorous interpretation of the qualitative data collected from our \total participants.
\cb{The process was iterative in practice, with recurring movement between phases as codes and themes were revisited and refined across multiple rounds (see Section~\ref{ta} for details).}

\begin{table}[ht]
\centering
\caption{Overview of the Thematic Analysis}
\label{TAsteps}
\begin{tabular}{p{3.3cm}|p{1.7cm}|p{5cm}|p{1.4cm}}
\hline
\rowcolor{gray!50}
\textbf{Step} & \textbf{Input} & \textbf{Activity} & \textbf{Output} \\
\hline
Data Familiarization & \textit{Transcripts} & Reading the transcripts multiple times, helping us make more accurate and insightful observations. & \textit{Initial notes} \\
\hline
Generation of Initial Codes & \textit{Initial notes}, \textit{transcripts} & Systematically identifying meaningful data segments (i.e., \textit{quotes}) and assigning them precise, descriptive \textit{codes}. & \textit{Initial codes} \\
\hline
Searching for Themes & \textit{Initial codes} & Searching for patterns in the data and grouping related codes into broader thematic categories. & \textit{Candidate themes}\\
\hline
Reviewing Themes & \textit{Candidate themes}, data & Adjusting \textit{candidate themes} by refining, discarding, or adding them to ensure coherence and consistency. & \textit{Refined themes} \\
\hline
Defining and Naming Themes & \textit{Refined themes} & Articulating and naming the essence of each theme & \textit{Final themes} \\
\hline
Producing the Report & \textit{Final themes}, \textit{quotes} & Creating the \textit{report} by using the \textit{final themes} and \textit{quotes}, linking them into a clear explanation supported by examples, and presenting the analysis. & Analytic narrative /report \\
\hline
\end{tabular}
\end{table}

\noindent\subsubsection{Data Familiarization}
We carefully read the \textit{transcripts} collected from our \total participants multiple times to develop a thorough understanding of their perspectives on SDs. \textit{Transcripts} are the verbatim, anonymized records of the semi-structured interviews conducted with our \total participants. These transcripts formed the primary data source of our study. During this phase, we also took informal \textit{initial notes} to capture early impressions and recurring observations.
For example, across multiple readings, we repeatedly encountered participants mentioning challenges related to unclear or vague security requirements, insufficient security expertise in development teams, and constraints imposed by outdated or legacy systems. These recurring observations highlighted potential areas of interest and laid the foundation for the structured coding process that followed.

\subsubsection{Generation of Initial Codes}
We systematically analyzed the highlighted \textit{initial notes}, along with the transcripts, to identify meaningful \textit{quotes} (i.e., data segments extracted from the transcripts). We documented these key aspects in an Excel sheet to ensure better organization and traceability of emerging patterns across participants and research questions. For instance, as depicted in Table~\ref{tab:quotes-codes}, we identified quotes that revealed underlying security challenges and their root causes, including ``Poor security requirements'' (P1), ``Lack of security expertise among developers'' (P7), and ``Vulnerabilities which may not be fixed due to older version of software... OS upgrade can be the only option but my legacy apps do not support upgraded version...'' (P14). These quotes highlighted recurring concerns, particularly deficiencies in security practices and systemic limitations caused by outdated infrastructure.

\begin{table}[ht]
\centering
\caption{Sample participant quotes and corresponding codes}
\begin{tabular}{p{5.5cm} | p{6.5cm}}
%\toprule
\specialrule{.1em}{0em}{0em}
\rowcolor{gray!50}
\textbf{Quote (Pid)} & \textbf{Code} \\
\specialrule{.1em}{0em}{0em}
``Poor security requirements.'' \textbf{ (P1)} & Inadequate security requirements \\ \hline
``Lack of security expertise among developers.'' \textbf{(P7)} & Insufficient developer security training \\ \hline
``Vulnerabilities which may not be fixed due to older version of software\ldots'' \textbf{(P14)} & Legacy system constraints \\
\bottomrule
\end{tabular}
\label{tab:quotes-codes}
\end{table}

\subsubsection{Identifying Initial Themes}
We analyzed the initial codes generated in the previous step to identify broader patterns and meaningful connections. Through this process, we grouped related codes into \textit{candidate themes}. For example, the codes “Inadequate security requirements” and “Insufficient developer security training” frequently co-occurred in responses referencing internal shortcomings in secure development practices. We tentatively grouped them under a candidate theme named \textit{Security Process Deficiencies}. In another case, the code “Legacy system constraints” was initially treated as an independent candidate theme, titled ``Technical Infrastructure Limitations,'' due to its specific technical nature.

\subsubsection{Reviewing Themes}

During the theme review phase, we revisited our \textit{candidate themes} in light of the full data. While \textit{Security Process Deficiencies} and \textit{Technical Infrastructure Limitations} were initially treated as distinct themes, further analysis revealed significant conceptual overlap. 
For example, participants often described the lack of formal security practices and outdated infrastructure as interrelated causes that together contributed to SD accumulation as part of answering RQ1 (see Section~\ref{objAndRQs}). Recognizing these inter-dependencies, we merged these candidate themes into a broader and more coherent refined theme, provisionally labeled \textit{Main Contributors to SDs}.

\subsubsection{Themes Definition \& Naming}
In this stage of the TA, we finalized the definition and naming of themes to ensure they accurately captured the essence of the patterns in the data. 
For example, the finalized themes for RQ1, along with their refined descriptions, are presented in Table~\ref{RQ1_analysis}. 
Following the same process, we categorized the entire dataset to ensure that the resulting themes conveyed a coherent and meaningful narrative around the research questions. In the previous step, we had refined related candidate themes, such as \textit{Security Process Deficiencies} and \textit{Technical Infrastructure Limitations}, into a broader theme provisionally labeled \textit{Main Contributors to SDs}. After revisiting and clarifying its scope, we finalized the theme as \emph{\textbf{root causes of SDs}}, as it better reflects the foundational and systemic nature of the factors described by participants.

\subsubsection{Final Report}
Although informal writing, such as research notes, began in the early phases, the final report was carefully prepared during the last stage of the TA. 
In this step, we systematically formulated and presented the results by directly addressing our research questions RQ1–RQ4, as elaborated in Sections~\ref{RQ1} through~\ref{RQ4}. 
The final report integrates the identified themes into a coherent analytic narrative, supported by illustrative quotes that authentically reflect participants’ perspectives and experiences.

\subsection{Data Analysis Execution}\label{ta}
\cb{The TA was conducted through an iterative coding and validation process involving two rounds of code validation and one round of theme validation.} 

\subsubsection{\cb{Iterative Coding Generation}}
\cb{The first author performed the initial open coding of all interview transcripts, identifying preliminary codes across all research questions (see Section~\ref{objAndRQs}). 
These codes underwent two validation rounds with the co-authors through consensus meetings, where codes were critically reviewed, refined, and reorganized based on team discussions, \minorrev{consistent with a reflexive thematic analysis approach}. 
For instance, in answering RQ1, and more precisely in answering the interview question 1.1 `How would you define an unresolved security vulnerability in any software?' (see the interview guide~\cite{guide}), Participant P5 stated: `Unresolved security vulnerabilities are when security requirements are missed, not properly implemented or not complying with client needs' was initially coded as `Detected but not remediated security issues'. After consensus discussion, we agreed to best fit it under `Inadequate or missing security requirements' code as it reflected missing or poorly implemented requirements rather than simply unresolved issues.}

\subsubsection{\cb{Iterative Theme Creation}}
\cb{Following the iterative coding process, we conducted a final theme validation round to group related codes into coherent themes, bringing the total number of qualitative refinement cycles to three across the entire thematic analysis process. For instance, in answering question 1.1 of the interview guide~\cite{guide}, the first author organized codes into candidate themes as follows: `Process and priority trade-offs' and `Deprioritized security issues' codes were grouped under the candidate theme `Process and priority trade-offs', while `Insufficient developer security training' and `Legacy system constraints' were grouped under `Technical/organizational constraint-related issues' candidate theme. However, during the theme refinement stage, the research team recognized significant conceptual overlap across these codes, as all reflected the deprioritization or neglect of security concerns due to limited resources (e.g., outdated libraries / software, and training allocation) or time pressure. Consequently, all four codes were consolidated into a single refined theme, Neglected security issues due to resource or time constraints, which more comprehensively captured how resource limitations and time pressures lead to the deprioritization and neglect of security issues in practice.
This iterative refinement process resulted in \cb{57} final themes across the four research questions: 19 themes for RQ1, 17 themes for RQ2, 12 themes for RQ3, and 9 themes for RQ4, which we further elaborate in Section~\ref{results}.}

\section{Research Findings}\label{results}
\label{sec:findings}

In the following, we report on the findings of our four research questions (see Section~\ref{objAndRQs}), providing insights, analysis, and key conclusions from our qualitative analysis on SDs. We articulate the themes we identified from the interview responses and present the findings for each theme, supported by direct quotations from participants' responses.

\subsection{\textbf{To what extent are software practitioners knowledgeable about SDs and their impact on software security? (RQ1)}}\label{RQ1}

In Table~\ref{RQ1_analysis}, we analyze software practitioners' understanding of SD according to six major themes: i) examining how software \textbf{\emph{definitions of SD}} provided by practitioners, ii) providing concrete \textbf{\emph{examples of SD}}, iii) identifying \textbf{\emph{root causes of SD}}, iv) illustrating \textbf{\emph{potential code vulnerabilities arising from SD}}, v) assessing the \textbf{\emph{time horizon of SD}}, whether they are viewed as short-term or long-term risks, and vi) evaluating the overall \textbf{\emph{impact of SD on software quality}}. \cb{For each theme, the table lists the key codes that emerged from our thematic analysis along with the number of participants who contributed to each code.}

\begin{table}[H]
    \renewcommand{\arraystretch}{1.25}
    \centering
    \caption{\cb{Themes of practitioner knowledge about SD (RQ1)}}
    \resizebox{\textwidth}{!}{
    \begin{tabular}{p{2.8cm} p{4.5cm} p{6.65cm} r}
        \hline
        \rowcolor{gray!50}
        \textbf{Theme} & \textbf{Description} & \textbf{Key Codes} & \textbf{Freq.} \\
        \hline  
        \multirow{1}{3cm}{Definition of SD}
            & \multirow{1}{4.5cm}{How practitioners conceptualize SD, ranging from latent system flaws to deferred resolution of known vulnerabilities.}
            & Known but unfixed issues & 10/22 \\
            \cline{3-4}
            & & Flaws from missed or poorly implemented requirements & 8/22 \\
            \cline{3-4}
            & & Neglected due to resource or time constraints & 8/22 \\
            \hline
        
        \multirow{1}{3cm}{Examples of SD}
            & \multirow{1}{4.5cm}{Concrete instances of unresolved security vulnerabilities encountered in real-world development and operations.}
            & Application-level vulnerabilities (e.g., SQLi, cookie issues, session management risks) & 8/22 \\
            \cline{3-4}
            & & Access control and authentication weaknesses (e.g., shared accounts, missing MFA, weak credentials) & 5/22 \\
            \cline{3-4}
            & & Outdated or misconfigured components (e.g., library vulnerabilities, insecure configurations, exposed secrets) & 5/22 \\
            \hline
        
        \multirow{1}{3cm}{Root Causes of SD}
            & \multirow{1}{4.5cm}{Organizational, technical, and contextual factors that prevent teams from resolving known vulnerabilities.}
            & Tight deadlines and time constraints & 12/22 \\
            \cline{3-4}
            & & Poor or missing security requirements & 11/22 \\
            \cline{3-4}
            & & Lack of security awareness, expertise, or oversight & 6/22 \\
            \cline{3-4}
            & & Poor communication and complex team dependencies & 3/22 \\
            \cline{3-4}
            & & Technical or infrastructure constraints (e.g., hardware limitations, domain-specific gaps) & 2/22 \\
            \hline
        
        \multirow{1}{3cm}{Potential Code Vulnerabilities Arising from SD}
            & \multirow{1}{4.5cm}{Types of security vulnerabilities practitioners anticipate from unresolved SDs, particularly at the code level.}
            & SQL/code injection (incl.\ prompt injection) & 11/22 \\
            \cline{3-4}
            & & Sensitive data disclosure and data manipulation & 6/22 \\
            \cline{3-4}
            & & Misconfigured access controls (RBAC) & 2/22 \\
            \cline{3-4}
            & & Other vulnerabilities (buffer overflows, DoS) & 2/22 \\
            \hline
        
        \multirow{1}{3cm}{Time Horizon of SDs (Short-term vs.\ Long-term)}
            & \multirow{1}{4.5cm}{How practitioners perceive the timeframe of risk associated with SDs, whether immediate, delayed, or both.}
            & Viewed predominantly as long-term risks & 19/22 \\
            \cline{3-4}
            & & Context-dependent (short-term in legacy or limited-maintenance systems) & 7/22 \\
            \hline
        
        \multirow{1}{3cm}{Impact of SD on Software Quality}
            & \multirow{1}{4.5cm}{How unresolved SDs affect software quality, including reliability, maintainability, and the CIA triad.}
            & Reduced reliability & 16/22 \\
            \cline{3-4}
            & & Increased maintenance overhead & 6/22 \\
            \cline{3-4}
            & & Confidentiality and integrity violations (data leaks, unauthorized modifications) & 4/22 \\
            \hline
    
    \end{tabular}
    }
    \label{RQ1_analysis}
\end{table}

\subsubsection{Definitions of SD}
This theme captures how practitioners conceptualize the meaning of SD. Their definitions reflect different interpretations, ranging from viewing SD as latent system flaws to understanding it as deferred resolution of known vulnerabilities. Clarifying how practitioners define SD is essential to understanding how they identify, manage, and prioritize such debts in real-world contexts.

\vspace{4pt}
\noindent\textbf{Findings.}
Participants conceptualized SDs primarily in two ways, though some definitions merged both perspectives. 
Several participants (P1, P2, P4, P7--P10, P19, P21, P22) defined SDs as latent \textit{system flaws} that could be exploited to make the software \textit{behave unexpectedly} (P1, P2), be compromised by attacks such as user impersonation (P18), or allow unauthorized access or manipulation of sensitive data (P7, P8). 
The presence of these vulnerabilities was also associated with delayed development timelines due to required fixes and testing (P21), and if left unaddressed, could lead to high remediation costs when exploited (P22).
For instance, P1 defined SD as:  \\
\begin{myquote}
    {\quoteFont{
    ``
    ... something that makes software not perform as expected.''
    }}
\end{myquote}

\cb{While this definition may superficially resemble a description of a general software bug, the context of P1's response clarifies that the unexpected behavior refers specifically to security exploitation, where an attacker leverages an unresolved vulnerability to cause the software to behave in unintended and potentially harmful ways, rather than an incidental software defect.}

SDs are also defined as intentional (P3, P5, P6, P9, P10, P12, P16) or unintentional (P18) unaddressed security vulnerabilities in software, possibly due to various constraints, 
such as lack of extensive software testing (P6, P20) and security deprioritization (P6, P10, P12), as quoted by P16: \\

\begin{myquote}
    {\quoteFont{
    ``...Vulnerabilities or problems which are intentionally ignored... possibility when you decide on whether you're delaying it, not fixing it at all, or taking it as a low priority thing where it might not exploit the system.''
    }}
\end{myquote}

\cor{Security-related and/or senior participants (e.g., P9, P10, P12, P16) emphasized SDs as intentional or deprioritized security vulnerabilities, while others (e.g., P1, P4, P7, P19) viewed SDs as latent flaws, suggesting that role and seniority shape how practitioners interpret the nature and origins of SDs.}

\noindent Overall, participants defined SDs either as exploit-prone system flaws that lead to unpredictable software behavior or as intentionally or unintentionally unresolved security vulnerabilities held back by several organizational and/or resource constraints\cor{, with security-related or senior participants more often highlighting intentionality and prioritization trade-offs, while others focused on technical flaws and potential exploitability.} 

\begin{takeawaybox}
\minorrev{\textbf{Definition of SD.}}
Practitioners define SD either as exploit-prone system flaws or as intentionally/unintentionally deferred vulnerabilities, with security-related and senior participants more often emphasizing intentionality and prioritization trade-offs.
\end{takeawaybox}

\subsubsection{Examples of SD}
\noindent This theme presents concrete examples shared by practitioners to illustrate what SDs look like in practice. These examples help clarify how unresolved security vulnerabilities manifest during real-world development and operations. The reported cases reflect a range of SD scenarios, from insecure configurations and missing requirements to persistent flaws that affected both system security and project timelines.

\vspace{4pt}
\noindent\textbf{Findings.}
Most participants (P1, P2, P4, P5, P7--P13, P14--P17) provided examples of unaddressed security vulnerabilities as incidents where systems ``{\quoteFont{cannot behave as expected}}'' or allow ``{\quoteFont{unauthorized access.}}''. Several recounted cases in which the flaws remained uncorrected. 
Some participants (P4, P7, P15) highlighted issues such as ``{\quoteFont{missing security requirements}}'' or \textit{``insecure configurations''}. P4 mentioned a shared account issue in his project:

\vspace{4pt}
\begin{myquote}
    {\quoteFont{``...Since the account is shared, we cannot track 
which support team member is responsible for actions, making it difficult to
 identify the source of an attack''}}
\end{myquote}
\vspace{4pt}

Further, P5 noted that vulnerabilities can result from missing controls, such as multi-factor authentication. P10 and P11 described SD instances involving outdated components and high-severity library vulnerabilities that blocked progress until resolved. P16 shared two examples: one involving a smart grid device vulnerable to remote control, and another where a desktop application allowed password extraction. P17 described difficulties enforcing the ``{\quoteFont{need to know}}'' principle in access control. P20 reported that front-end timeouts in the system caused delayed responses, leading to session management risks.
P21 shared an incident in which inadequate firewall protections and exposed keys in a wallet system led to financial loss and significantly slowed project progress, and P22 mentioned the case of accepting weak credentials as an example of SD.

These examples highlight how technical oversights and organizational constraints often interact to leave systems exposed to persistent, unresolved security risks.

\cor{Security-focused and/or more senior participants (e.g., P10, P15, P16, P17) tended to report \cb{more complex and system-level SD instances, such as outdated components, high-severity library vulnerabilities, and access control flaws requiring architectural changes, as opposed to commonly known issues such as missing Multi-Factor Authentication (MFA) or weak credentials.}} 
\cb{This indicates that the role and experience of the participants may influence the depth and nature of the SD examples shared.}

\begin{takeawaybox}
\minorrev{\textbf{Examples of SD.}}
Reported SD examples range from access control and credential flaws to outdated components and \emph{application-level vulnerabilities}, with role and experience shaping the depth and nature of examples shared.
\end{takeawaybox}

\vspace{4pt}
\subsubsection{Root Causes of SD}
\noindent This theme explores the reasons why SDs remain unaddressed in practice. It captures the organizational, technical, and contextual factors that prevent teams from resolving known vulnerabilities. Understanding these root causes is key to identifying what constraints or trade-offs lead to postponed security work.

\vspace{4pt}
\noindent\textbf{Findings.}
Participants identified a variety of factors contributing to unaddressed security vulnerabilities, \cb{with some participants sharing more than one cause of SD}.
The most prominent cause of SD was time constraint (P2--P4, P9, P11--P13, P15, P19--P22), particularly due to the complexity of implementing security requirements under tight deadlines (e.g., ``...unfortunately...we have less time for security...'', quoted P15). This often led to prioritizing delivery over testing and code review. The second most common factor was \emph{poor or missing security requirements}, often caused by prioritizing functional requirements over security considerations (P1, P3, P4, P5, P10, P12, P13, P15, \cor{P16}, P17, P18).
\cb{For example, P10 said ``...fixing security issues requires downtime, and since it is not directly exposed to the internet, it is considered low priority...'', through which security updates were deprioritized in favor of keeping the system running.}

Some participants mentioned other contributing factors such as lack of technical expertise \cb{such as P6, P7, P9, P15, P16, and P22 (e.g., ``some programmers modify existing code without fully understanding security implications, leading to vulnerabilities'', said P7)}, poor communication (P5, P6), lack of attention to detail (P5), non-compliance with industry standards (P5), and complex dependencies between teams (P4, P13). 
\cb{For example, P4 said ``shared account access...since the account is shared, we cannot track which support team member is responsible for actions...'', through which team dependencies left vulnerabilities unresolved.}
Additionally, some participants pointed to lack of supervision and insufficient review for junior developers (P7), inadequate domain-specific monitoring (e.g., in blockchain systems) (\cb{e.g., ``very poor monitoring by blockchain people'', quoted P21}), \cor{and embedded system limitations such as  the inability to perform full encryption due to hardware constraints (\cb{e.g., ``they cannot always encrypt all the traffic at high speeds'', quoted} P16) as further root causes.}

\cor{Overall, time constraints and poor security requirements were common causes across roles (P1--P5, P9--P13, P15, P17--P22). While junior to mid-level and/or non-security participants (e.g., P5, P6, P9, P19, P22) more often cited lack of expertise or oversight, senior or security-focused roles (e.g., P13, P15--P17) pointed to systemic issues like complex team dependencies, code complexity, and embedded system constraints. This suggests that the role and experience influence how root causes are identified.}

These responses suggest that SDs are primarily driven by time pressures and the prioritization of functionality over security, with trade-offs made to meet business goals and deadlines at the expense of secure practices.
\cb{SD cause findings are consistent with existing technical debt literature, where time constraints and unclear requirements are well-established drivers of debt accumulation~\cite{kruke2024defining, rindell2019managing, coetzer2024managing, huopio2020quest}. However, in the security context specifically, these pressures carry heightened consequences, leaving known vulnerabilities unpatched rather than merely deferring code quality improvements, with direct exploitation risks for software systems.}

\cb{The behavioral consequences of these root causes, particularly how practitioners respond to deadline pressure through prioritization decisions and security trade-offs, are examined in depth under RQ2.}
\cb{This pattern of security deprioritization in favor of functional delivery emerges as a cross-cutting theme throughout our findings and is not restated in subsequent sections to avoid repetition; rather, it is acknowledged here as the overarching organizational dynamic underlying many of the behaviors and challenges reported across the remaining RQs.}

\begin{takeawaybox}
\minorrev{\textbf{Root Causes of SD.}}
Time constraints and poor security requirements are the dominant root causes of SDs, with senior and security-focused participants more often pointing to systemic issues such as complex team dependencies and infrastructure constraints.
\end{takeawaybox}

\subsubsection{Potential Code Vulnerabilities Arising from SD}
\noindent This theme focuses on the specific types of security vulnerabilities practitioners believe are likely to emerge from unresolved SDs. These anticipated issues, particularly at the code level, help illustrate the practical consequences of leaving SDs unaddressed.

\vspace{4pt}
\noindent\textbf{Findings.}
Most of the participants (P1, P2, P4, P5, P16--P22) stated that code injection attacks, notably SQLi, are the most likely \emph{\textbf{potential code vulnerabilities arising from SDs}}. P5 shared an example with code injection.

\vspace{4pt}
\begin{myquote}
    {\quoteFont{
    ``Code injection to compromise code such as SQLi... create malicious payloads for dataset manipulation''
    }}
\end{myquote}
\vspace{4pt}
\cb{This finding is consistent with prior research and industry reports identifying code injection attacks, notably SQLi as one of the most prevalent types of software security weaknesses~\cite{owasp_top10_2021, riskhan2025major, ndebugre2025comprehensive,edbert2023exploring,tony2022conversational,diaz2024can}.}

Some participants (e.g., P20, P21) emphasized evolving forms such as prompt injection particularly relevant to AI systems and malicious payloads that could expose models or sensitive data. These concerns reflect a shift toward not only web-based vulnerabilities but also domain-specific and AI-integrated risks. While P22 reported applying input validation to mitigate a real injection incident, others stressed the insufficiency of such controls without systemic security improvements.
After code injection, the most likely code vulnerabilities are sensitive data disclosure (P3, P8, P9, P16) and data manipulation (P4, P5), followed by Role-Based Access Control (RBAC) vulnerabilities (P1, P8), buffer overflows (P16), and Denial of Service (DoS) attacks (P4).
P3 quoted: \\

\begin{myquote}
    {\quoteFont{
    ``Sensitive data disclosure such as exposure of unwanted services to public, or exposure of passwords or private user info in URLs, web page scripts)''
    }}
\end{myquote}
\vspace{7pt}
and P20 added: \\
\begin{myquote}
    {\quoteFont{
    “any code injection...even prompt injection on our own GPT.”
    }}
\end{myquote} 

\cor{Code injection, especially SQLi, was cited as a key risk from SD by several participants, including P5, P16, and P20, with P22 describing a real incident mitigated through input validation. While some security-focused (e.g., P16, P17) and senior participants (e.g., P14--P17) framed these threats using the CIA triad or linked them to broader risks, similar concerns were raised across various roles and experience levels. Therefore, we notice that no strong role-based or seniority-based pattern emerged, indicating that awareness of code-level attacks, especially code injection attacks, is common among software practitioners with all levels and domains of expertise.}

\cb{The vulnerabilities listed here are not intended to be exhaustive but reflect specifically those anticipated by study participants based on their real-world experience with SDs, and are therefore reported as participant-driven findings rather than a comprehensive taxonomy of security vulnerabilities.}
Our findings conclude that code injection vulnerabilities, followed by data disclosure and manipulation, are most likely to arise due to SDs, representing a primary threat to the confidentiality and integrity aspects of the CIA triad, while DoS attacks pose a risk to availability.

\begin{takeawaybox}
\minorrev{\textbf{Potential Code Vulnerabilities Arising from SD.}}
Code injection (especially SQLi and prompt injection) is the most anticipated vulnerability from SDs, followed by \emph{sensitive data disclosure and data manipulation}, threatening the confidentiality and integrity aspects of the CIA triad.
\end{takeawaybox}

\subsubsection{Time Horizon of SDs (Short-Term versus Long-Term Risks)}
\noindent This theme reflects how practitioners perceive the timeframe of risk associated with SDs, whether their impact is immediate, delayed, or both. Clarifying the \emph{\textbf{time horizon of SDs}} helps reveal how the latter are prioritized and managed across different development contexts.

\vspace{4pt}
\noindent\textbf{Findings.}
There was strong consensus among participants (P1–\cor{P19}) that unaddressed security vulnerabilities primarily lead to long-term risks. 
P18 and P19 acknowledged the presence of both short-term and long-term risks.
Several participants noted that while the immediate impact of these vulnerabilities may be minimal, they can ``{\quoteFont{damage the software eventually}}'' (P1, P6, P9, P10, P15, P16), and often go unnoticed until they accumulate into significant issues (P2, P10, P15). 

Although most participants viewed SDs as long-term threats to software, some (P3, P8, P12, P16, P17, P20, P21) emphasized that SDs can also pose short-term risks, depending on the software’s nature, intended purpose, and functionality. For example, P3 mentioned a case involving software developed for exceptional, short-term circumstances.
For example, P7 illustrated this as:

\vspace{4pt}
\begin{myquote}
    {\quoteFont{
    ``Fixing vulnerabilities after they are discovered often requires significant effort, and in some cases, a complete redesign of system e.g., I worked with financial company where a security breach led to a system-wide investigation and an expensive overhaul.''
    }}
\end{myquote}
\vspace{4pt}

Some participants (P8, P9, P12, P14, P15, \cor{P18, P19}) emphasized that when security measures are left unaddressed, the resulting issues evolve into entrenched risks that become more difficult (and more expensive) to resolve. 
P16 noted that the time horizon of SDs may depend on the level of system exposure: while publicly accessible applications might be exploited quickly, even secured environments can experience long-term consequences. P17 added that if these issues remain unresolved, they ``{\quoteFont{could have long-term consequences}}.''

\cor{
Security experts (P10, P12--P17) and senior participants (P7, P14--P18) identify SDs as both short-term and long-term risks, with some seniors emphasizing remediation costs. Less senior or non-security participants (e.g., P1--P3, P6, P8, P19, P21) mainly see SD as a long-term issue. P16 notes system exposure affects risk timing. Overall, greater expertise and seniority show a context-aware perception of SD risks regarding timeframe, while less experienced and non-security roles typically view these risks in broader, longer-term terms.
}

These responses suggest that unimplemented security measures, while sometimes deprioritized in the short term, create persistent risks that compromise system integrity and eventually demand costly remediation.

\begin{takeawaybox}
\minorrev{\textbf{Time Horizon of SD.}}
SDs are predominantly viewed as long-term risks, though security experts and senior participants recognize context-dependent short-term risks as well, reflecting a more nuanced, experience-driven perception of SD timeframes.
\end{takeawaybox}

\subsubsection{Impact of SD on Software Quality}
\noindent This theme captures practitioners’ perspectives on how unresolved SDs affect software quality, a critical concern in both engineering and business contexts. Understanding this impact is essential because reliability and maintainability directly influence user trust, system longevity, and operational costs.

\vspace{4pt}
\noindent\textbf{Findings.}
All participants (P1--P22) agreed that unaddressed security vulnerabilities significantly impact software reliability and maintainability. Many participants (P1, P2, P3, P4--P6, P9, P12, P13, P14, P15, P18--P22) explicitly stated that reliability suffers, with P5 noting:

\vspace{4pt}
\begin{myquote}
    {\quoteFont{``It can be compromised anytime… it also might crash any time.''}}
\end{myquote}
\vspace{4pt}

Six participants (P5, P6, P9, P10, P11, P12) claimed that maintainability is also affected by SDs:

\vspace{4pt}
\begin{myquote}
    {\quoteFont{``making future security efforts harder... over time, systems become difficult to maintain due to unclear access permissions \& undocumented configurations.''}}
\end{myquote}
\vspace{4pt}

Several participants (P3, P9, P14, P16) linked security vulnerabilities to confidentiality-related issues, notably data leaks, compliance risks, and reputational damage. P3 emphasized, ``{\quoteFont{Data leaks… may impact the company's public credibility and reputation.}''
P9 added concerns about ``{\quoteFont{inconsistent processes and increased maintenance overhead.}''
P16 also links SDs with integrity, stating, as a context example, banking applications where ``{\quoteFont{someone can gain access to user data or modify banking details or perform unauthorized transactions...}'' }}
P8 warned that insecure foundations create a cycle where ``{\quoteFont{insecurities keep accumulating,}}'' and suggested that ``{\quoteFont{Instead of just adding new features, I recommended rebuilding the system with stronger security measures.}}''

Others participants (P10, P11) pointed out that vulnerabilities increase SD, making updates harder due to ``{\quoteFont{unclear access permissions and undocumented configurations.}}''
P15 observed that after security breaches, some companies ``{\quoteFont{just put it offline and outsource it... they just wanted to get it replaced as soon as possible.}''
P16 also noted that the impact severity can vary depending on the application, with banking vulnerabilities enabling ``{\quoteFont{unauthorized transactions}}'' while website defacement is mostly reputational. P13 stressed that the impact sometimes depends on ``\quoteFont{the likelihood of this to be exploited.}''

\cor{While all participants agree on the impact of SDs on software maintainability and especially reliability, most security-role participants (P10, P12--P16) discuss specific impacts of SDs on reliability, maintainability, confidentiality, and integrity with concrete examples. This suggests that seniors and security-related participants have a deeper, more concrete understanding of SD's impact on software quality.}
}

Overall, unaddressed security flaws create cascading issues, leading to unreliable software, costly maintenance, and potential long-term business risks, along with breaches of software confidentiality, integrity, and availability.

\begin{takeawaybox}
\minorrev{\textbf{Impact of SD on Software Quality.}}
Participant background influenced SD perception across RQ1 themes. Senior practitioners identified broader systemic causes and risks, while junior practitioners focused on technical issues and time constraints. Across regions, delivery pressures and intentional deprioritization were common drivers, and participants consistently linked SDs to reduced reliability, security risks, and increased remediation costs.
\end{takeawaybox}

\subsubsection*{Cross Background, Experience, and Regional Pattern Analysis}
\noindent
%Table~\ref{RQ1_analysis} reports findings across all RQ1 themes.
Examining RQ1 findings across all RQ1 themes against participants' backgrounds reveals several patterns. Regarding \textit{experience level}, junior and mid-level participants (1--5 years) predominantly framed SDs as latent technical flaws and cited time constraints as the main root cause, while senior and highly senior participants more frequently identified systemic causes such as embedded constraints and complex team dependencies. A seniority gradient also emerged in SD examples, where senior and highly senior participants (e.g., P15--P17) reported deeper systemic issues such as outdated components and access control flaws, while junior and mid-level participants (e.g., P4, P5, P22) described more commonly known issues such as missing MFA or weak credentials. The dual short- and long-term risk perception was similarly concentrated among highly senior participants (P16--P18). For \textit{code vulnerabilities}, no background pattern emerged for code injection awareness, indicating broad familiarity across all experience levels and regions. Regarding \textit{region}, Asia-based participants (PK, IR) were notably concentrated among those citing intentional deprioritization and poor security requirements, while North America-based participants more often reported time constraints. Notably, the \textit{impact of SD on software quality}, particularly reduced reliability, was the only theme where no background-related divergence was observed, with unanimous agreement across all experience levels, roles, and regions, consistent with the overall conclusion that all participants, regardless of background, recognize SD as a threat to software reliability with cascading effects on maintainability, confidentiality, and integrity.
\vspace{.3cm}

\begin{summarybox}[RQ1 Summary]
    \minorrev{SDs arise from unresolved security vulnerabilities and are primarily driven by technical, resource, and organizational constraints. Practitioners viewed SDs as a source of long-term risk that can compromise system security, reliability, and maintainability, highlighting the need to balance security priorities with business and development pressures.}
\end{summarybox}

 \subsection{\textbf{How do software practitioners behave towards SDs in software development? (RQ2) }}\label{RQ2}

\noindent We analyze how practitioners handle and address SDs. Our analysis focuses on \cb{four} themes, as shown in Table~\ref{RQ2_analysis}: \emph{addressing SDs}, \cb{\emph{security-deadline trade-offs and prioritization}}, \emph{preventing SD accumulation}, and \emph{persistent security vulnerabilities}. 
\cb{For each theme, the table shows the key codes and the number of participants who contributed to each.}

%% NEW TABLE %%
\begin{table}[H]
    \renewcommand{\arraystretch}{1.25}
    \centering
    \caption{\cb{Themes of practitioner behavior toward SDs (RQ2)}}
    \resizebox{\textwidth}{!}{
    \begin{tabular}{p{2.8cm} p{4.5cm} p{6.65cm} r}
    \hline
    \rowcolor{gray!50}
    \textbf{Theme} & \textbf{Description} & \textbf{Key Codes} & \textbf{Freq.} \\
    \hline   
    \multirow{1}{3cm}{Addressing Identified Vulnerabilities}
        & \multirow{1}{4.5cm}{How practitioners handle identified SDs, including who takes responsibility and whether resolution is direct or delegated.}
        & Immediate remediation & 4/22 \\
        \cline{3-4}
        & & Severity assessment and recommendation & 2/22 \\
        \cline{3-4}
        & & Delegation and escalation to dedicated teams & 4/22 \\
        \hline
    \multirow{1}{3cm}{\cb{Security-Deadline Trade-offs and Prioritization}}
        & \multirow{1}{4.5cm}{How teams prioritize SD resolution under strict time constraints, and what trade-offs are involved in balancing risk with delivery demands.}
        & All vulnerabilities addressed (no prioritization) & 7/22 \\
        \cline{3-4}
        & & Severity-based prioritization & \cb{18}/22 \\
        \cline{3-4}
        & & Client or context-driven prioritization & 4/22 \\
        \cline{3-4}
        & & Effort-based prioritization & 3/22 \\
        &  &\cb{Extra effort or resource allocation (e.g., overtime, additional staff)} & \cb{8/22} \\
        \cline{3-4}
        & & \cb{Deadline extensions} & \cb{4/22} \\
        \hline
    \multirow{1}{3cm}{Preventing Vulnerability Accumulation}
        & \multirow{1}{4.5cm}{Behavioral strategies practitioners use to prevent SD accumulation, including organizational practices and lightweight technical routines adopted as regular development habits.}
        & Communication and internal meetings & 6/22 \\
        \cline{3-4}
        & & Regular tracking and backlog management & 6/22 \\
        \cline{3-4}
        & & Continuous resolution of issues & 3/22 \\
        \cline{3-4}
        & & Dedicated security time allocation & 3/22 \\
        \cline{3-4}
        & & Use of automated tools & 6/22 \\
        \hline
    \multirow{1}{3cm}{Persistent Security Vulnerabilities}
        & \multirow{1}{4.5cm}{Cases where security vulnerabilities remain unresolved for extended periods, and the factors that lead to long-lived SDs.}
        & No long-standing security issues & 8/22 \\
        \cline{3-4}
        & & Technical constraints preventing resolution (e.g., misconfigurations, implementation difficulties) & 8/22 \\
        \cline{3-4}
        & & Emerging threats and evolving vulnerabilities (e.g., database or wallet hacks) & 4/22 \\
        \hline 
    \end{tabular}%
    }
    \label{RQ2_analysis}
\end{table}

\subsubsection{Addressing SDs}
\noindent This theme describes how practitioners handle identified SDs in their projects. It is important to understand who takes responsibility for resolving \cor{security} vulnerabilities, how actions are taken, and whether resolutions are direct or delegated.

\vspace{4pt}
\noindent\textbf{Findings.}
Only four participants (P1, P2, P8, P9) \cb{reported addressing SDs in projects assigned to them by taking corrective action when issues were identified. This suggests that addressing SDs was not commonly observed among participants.} For instance, P2 updated the implementation to ensure data non-disclosure: ``{\quoteFont{...forgot about data non-disclosure in our implementation and addressed it by updating code}}''. Participant P8 mentioned a case in which they worked on improving API security and password validation, and P9 gave an example of fixing SDs that involves the enforcement of uniform security procedures to prevent data leaks. 

Although most of the participants did not share their experience in addressing concrete SD examples, P7 provided a comprehensive illustration of how to address SD in practice, saying: \\

\begin{myquote}
    {\quoteFont{
    ``I first assess the severity of vulnerability. If it is critical, I address it immediately. If it is minor, I may schedule it for a later fix, depending on deadline.''
    }}
\end{myquote}

\cor{Similarly, P10 added: \\
\begin{myquote} 
    {\quoteFont{``my role is to identify vulnerabilities and provide recommendations for fixing them. I do not directly fix them...''}}
\end{myquote}} 

Some other participants (P5, P11, P15, P16) stated that such tasks fall outside their role scope, e.g., ``{\quoteFont{...not in the projects assigned to me}}'' (P5), and that they typically report the issues to dedicated teams. For instance, P11 mentioned that he does ``{\quoteFont{not directly identify vulnerabilities. Vulnerabilities are detected and reported automatically by internal software and third-party vendors}}.''

\cor{While seniors and security-focused participants (e.g, P7, P10) typically assess SDs and their criticality to software quality, often delegating remediation to other teams, non-senior and non-security roles (e.g., P1, P2, P8, P9) were more likely to perform hands-on remediation, suggesting a division of responsibility by role type.}

Overall, participants approached SDs in nuanced ways. \cor{On one hand, juniors mostly undertook} immediate corrective actions, such as updating implementations and enforcing security measures, while \cor{senior or security role-based participants often examine SDs and provide appropriate recommendations to fix them.} 
\cor{On the other hand, some other participants, regardless of their roles and experience, reported the issues to relevant teams or stated that addressing them was beyond their job responsibilities. However, these participants were uncertain if the responsible teams addressed these issues promptly or if there were delays, as there were no notification mechanisms in place.}

\begin{takeawaybox}
\minorrev{\textbf{Addressing SD.}}
Junior practitioners tend to perform immediate hands-on remediation, while senior and security-focused participants assess severity and delegate fixes, with some participants reporting SDs to other teams without certainty of timely resolution.
\end{takeawaybox}

\subsubsection{\cb{Security-Deadline Trade-offs and Prioritization}}
\noindent \cb{This theme explores how teams prioritize SD resolution under strict time constraints, and how practitioners manage the resulting conflict between security needs and delivery demands.}

\vspace{4pt}
\noindent\textbf{Findings.}
The prioritization of addressing SDs in projects with tight deadlines varies across teams and depends on multiple factors. Some participants (P1, P2, P7–P9, P18, P20) believe all SDs should be addressed equally, with no prioritization (e.g., {\quoteFont{``all important, no prioritization''}}, said P1), and insist that SDs take precedence over deadlines, which can be postponed if necessary (P1, P8, P9).

However, most participants agree that prioritization should be based on severity and impact on software reliability (P3, P10–P17, P19), along with other factors such as task complexity (P5) and overall importance (P4). Some (P6, P15, P16) prioritize SDs primarily based on client needs, which often influence whether vulnerabilities are addressed immediately or postponed, as P6 noted: {\quoteFont{“...it's all about the client needs."}}. For example, one participant (P16) explained:

\vspace{4pt}
\begin{myquote}
    {\quoteFont{
    ``I worked with clients in the US who were building applications to sell. They had to decide whether to fix the issue right away or delay it, depending on their priorities.''
    }}
\end{myquote}
\vspace{4pt}

Other participants (P12, P22) shared that prioritization may depend on project managers or specific contexts, such as hospital data access \cb{(e..g, ``...escalate them to the manager for a final decision'': P12)}.
P21, who is working in a startup, emphasized strict adherence to deadlines, relying on rewards, overtime, and dedicated testing teams to ensure timely delivery without compromising necessary fixes.
\cor{Senior and security-focused participants (e.g., P10, P12--P17, P14--P18) prioritized SDs based on their severity, impact on software reliability, or even client needs, despite the expectation that technical risk should come first, particularly from a senior security participant. In contrast, those favoring no prioritization and treating all SDs equally (P1, P2, P9) were neither senior nor in security roles, suggesting that experience and specialization influence prioritization approaches.}

Overall, while some, \cor{mostly juniors or those with non security-related roles} treat all SDs equally, \cor{most seniors and participants with security-based roles} prioritize SD severity and \cor{their} impact on \cor{software} reliability, balancing security efforts with tight deadlines and business needs.

\cb{Beyond how SDs are ranked, practitioners also described the concrete mechanisms they use to manage the security-delivery conflict in practice.}

Participants suggested various ways to balance security and strict delivery deadlines: working overtime (P1, P3, P9, P20), possibly extending deadlines if security is not fully met; delivering software with less time-consuming security tasks completed (P5); dividing tasks across teams (P6); addressing the first identified issues (P8, P10, P11); and focusing on the most critical security tasks (P12, \cor{P16,} P17, P19, P21). Some also mentioned compensating with extra hours or involving additional team members (P18), \cb{maintaining close collaboration with DevOps through CI/CD pipelines and automated security checks, postponing deadlines when necessary to address client-driven security concerns} (P20), or conducting regular meetings and code reviews to avoid critical delays (P22).
\cor{Participants in security or senior roles (e.g., P10, P12, P16--P18) tended to prioritize based on risk (e.g., addressing the most critical issues) and use strategic coordination (e.g., involving teams or seeking extra support) to balance security and deadlines. In contrast, most junior participants (e.g., P1, P3, P9, P20) tended to rely on overtime, extend deadlines, or prioritize less time-consuming SDs (e.g., P5) or the first identified ones (e.g., P8, P10, P11). The pattern shows that participants with senior and/or security roles rely on risk-based and coordinated strategies, while junior participants tend to use overtime or quick fixes. This difference highlights the influence of experience on balancing security and deadlines.
}
\cb{Participants described different approaches to managing security tasks under tight timelines.} While some participants prioritize security even at the expense of deadlines (e.g., accepting overtime to ensure issues are resolved), others focus more on timely delivery, potentially compromising the depth or thoroughness of the implemented security measures. \cor{Notably, senior and security-focused participants tend to manage the security and deadline trade-off through risk-based prioritization and strategic coordination, whereas junior participants more often rely on overtime or quick fixes, highlighting how experience and security expertise shape approaches to balancing security with delivery pressure.}

\begin{takeawaybox}
\minorrev{\textbf{Security-Deadline Trade-offs and Prioritization.}}
Senior and security-focused participants balance deadlines through risk-based prioritization and strategic coordination, while junior participants more often treat all SDs equally and rely on overtime or quick fixes, highlighting how experience shapes the management of security and delivery trade-offs.
\end{takeawaybox}

\subsubsection{Preventing Vulnerability Accumulation}
\noindent This theme captures the strategies practitioners use to \cb{prevent SD accumulation, including both organizational practices and lightweight technical routines adopted as regular development habits.}

\vspace{4pt}
\noindent\textbf{Findings.}
Participants adopt different approaches to prevent the accumulation of SDs over time. Communication and internal meetings (P1, P5, P6, P8, P9, P19) help maintain awareness, while others (P5, P12) recommend allocating more time for security tasks and, when possible, hiring additional technical staff. Some participants focus on resolving issues immediately upon identification (P3, P6, P7), or tracking them systematically via backlogs, as P4 stated: ``{\quoteFont{...maintaining a backlog, ensuring SDs are tracked and addressed systematically.}}''

Some participants P10--P13 believe that regular software updates, access control policies, and automated tracking tools (P11, P12, P13, P20) further help prevent SD accumulation. Other strategies include code review routines and Git-based issue tracking processes (P22). Participants also mentioned using security-oriented CI/CD pipelines, having dedicated security teams (2–3 people) for compliance checks (P21), and practices such as automated benchmark testing, internal security discussions, and extensive testing and documentation (P18, P20).

Findings show that participants use a mix of proactive and reactive strategies to prevent SD accumulation, including communication, adoption of security measures, and both manual and automated tracking. This underscores the importance of organizational practices (e.g., communication, time allocation, security backlogs, implementation of security policies and Agile methods) and technical solutions (e.g., automated tracking tools, vulnerability scanners such as
\textit{SonarCloud}) in effectively managing SDs over time.

\cb{Although automated tools are also examined under RQ3 as part of SD mitigation practices (see Section~\ref{RQ3}), their appearance in RQ2 reflects a distinct finding: when discussing their behavioral practices toward SDs, practitioners consistently framed automated tools as routine prevention habits rather than deliberate mitigation responses. This distinction is analytically meaningful and is therefore reported under both RQs.}

\begin{takeawaybox}
\minorrev{\textbf{Preventing Vulnerability Accumulation.}}
Participants use a mix of proactive and reactive strategies to prevent SD accumulation, combining organizational practices (e.g., communication, backlog management, dedicated security time) with technical solutions (e.g., automated tracking tools, CI/CD pipelines).
\end{takeawaybox}

% \subsubsection{Balancing Deadlines and Security}
% \noindent This theme examines how practitioners manage the conflict between delivery timelines and the need to address security concerns. Understanding such a balance helps identify what compromises are made and how teams attempt to minimize risk without delaying releases.

% \vspace{4pt}

\subsubsection{Persistent Security Vulnerabilities}
\noindent This theme focuses on cases where security vulnerabilities remain unresolved for extended periods. These cases highlight the factors that lead to long-lived SDs, such as technical complexity, organizational gaps, or resource limitations.

\vspace{4pt}
\noindent\textbf{Findings.}
Nearly half of the participants (P2--P4, P6, P8, P9, P13, P22) reported no significant long-standing security issues in their projects. However, several others (P1, P5, P11, P14, P19–P21) recounted cases where certain bugs persisted due to various challenges. These included resource constraints such as memory limits (P1), complex misconfiguration issues (P10), and technical challenges, such as implementing security logging and protection mechanisms (P16), for example, a login bypass in a desktop application. Additional obstacles included implementation difficulties (P5, P15, P17), authentication-related issues (P18), and delays in resolving major incidents such as database or wallet hacks (P20, P21), as illustrated by P5:

\vspace{4pt}
\begin{myquote}
    {\quoteFont{
    ``The OTP implementation was first delayed, then not properly implemented... not many people were available for testing the functionality.''
    }}
\end{myquote}
\vspace{4pt}

\cor{Persistent security issues were more often reported by senior and security-focused participants (e.g., P14--P18), who tended to describe complex, systemic challenges such as misconfigurations, authentication flaws, or integration limitations. In contrast, those reporting no major issues (e.g., P2--P4, P6, P8, P9) were generally not in senior or security roles, suggesting that experience and role depth may lead to greater awareness or exposure to long-standing security risks often overlooked in less technical or less exposed roles.}

Findings highlight that implementation challenges, including misconfigurations and difficulties in integrating security mechanisms, are key obstacles to addressing SDs, suggesting the need for specialized expertise and greater team collaboration.

\begin{takeawaybox}
\minorrev{\textbf{Persistent Security Vulnerabilities.}}
Persistent SDs are more often reported by senior and security-focused participants, who describe complex systemic challenges such as misconfigurations and authentication flaws, suggesting that experience leads to greater awareness of long-standing security risks.
\end{takeawaybox}

\subsubsection*{Cross Background, Experience, and Regional Pattern Analysis}
\noindent
%Table~\ref{RQ2_analysis} maps RQ2 findings against participant background,
Analyzing RQ2 findings against participant background reveals consistent experience- and region-related patterns across its \cb{four} themes. Regarding \textit{addressing identified vulnerabilities}, junior and mid-level participants (P1, P2, P8, P9) predominantly performed immediate hands-on remediation, while senior and highly senior participants (P7, P15, P16) more often conducted severity assessments or delegated fixes to dedicated teams, reflecting a division of responsibility shaped by experience and role depth. Concerning \textit{prioritization under tight deadlines}, the no-prioritization approach was concentrated among junior and mid-level participants (e.g., P1, P2, P8, P9), whereas severity-based prioritization was almost exclusively reported by mid-level to highly senior participants (P10, P13--P17, P19), with Asia-based participants most frequently applying client- or context-driven prioritization. For \textit{preventing vulnerability accumulation}, junior participants favored communication and immediate resolution, while mid-level to highly senior participants more often adopted systematic approaches such as backlog management and automated tools, with North America-based participants notably concentrated among those using automated tracking solutions. Regarding \textit{balancing deadlines and security}, junior participants predominantly relied on overtime or deadline extensions, whereas senior and highly senior participants applied risk-based prioritization and strategic coordination. Finally, for \textit{persistent security vulnerabilities}, the absence of long-standing issues was almost exclusively reported by junior and mid-level participants, 
while complex systemic challenges such as misconfigurations and authentication flaws were reported solely by senior and highly senior participants (P14--P18).

%\begin{takeawaybox}
%Participant background influenced behavior toward SDs across RQ2 themes. Junior and mid-level practitioners favored direct remediation, while senior practitioners adopted risk-based prioritization and coordination strategies. Regional differences were also observed, with Asia-based participants emphasizing client-driven prioritization and North America-based participants reporting greater use of automation and preventive practices.
%\end{takeawaybox}

\vspace{3pt}
\begin{summarybox}[RQ2 Summary]
    \minorrev{Practitioners manage SDs through a combination of proactive and reactive approaches, typically prioritizing remediation based on risk and impact. Management practices are shaped by practitioner expertise, organizational constraints, and the need to balance security requirements with delivery pressures.}

    %\cb{Overall, practitioners’ behavior toward SDs varies significantly by role and experience, particularly in how they perceive, prioritize, and respond to security issues in development contexts.} Participants manage SDs through a combination of proactive and reactive strategies, often prioritizing based on severity and impact. \cb{Proactive strategies included communication and internal meetings, regular backlog management, and use of automated tools such as CI/CD pipelines and SonarCloud.} 
    %\cor{With more complex software security challenges typically recognized and addressed by experts and/or more experienced practitioners, compared to juniors and/or those occupying non-security-specific roles}. Addressing SDs requires balancing deadlines with security needs, relying on organizational practices, and overcoming technical challenges through team coordination and specialized expertise.
    %Finally, regional background also influenced approaches, with Asia-based participants more frequently applying client-driven prioritization and North America-based participants more often adopting automated prevention strategies.
    %\cb{Specifically, immediate remediation was performed by only 4/22 practitioners, severity-based prioritization was the most common approach (10/22), and persistent security vulnerabilities were reported by several participants, most often involving misconfigurations and authentication flaws. Communication and backlog management (6/22 each) were the most frequently adopted prevention strategies.}
\end{summarybox}

\subsection{\textbf{How do software practitioners mitigate SDs in software? 
(RQ3) }}\label{RQ3} 

To answer RQ3, as illustrated in the three themes defined in Table~\ref{RQ3_analysis}, we analyze how practitioners mitigate SDs.
We also investigate the role of AI, when adopted in security, as well as the challenges practitioners encounter in SD mitigation. \cb{For each theme, the table shows the key codes and the number of participants who contributed to each.}

%% NEW TABLE %%
\begin{table}[H]
    \renewcommand{\arraystretch}{1.25}
    \centering
    \caption{\cb{Themes of SD mitigation practices and challenges (RQ3)}}
    \resizebox{\textwidth}{!}{
    \begin{tabular}{p{2.6cm} p{4.5cm} p{7.03cm} r}
    \hline
    \rowcolor{gray!50}
    \textbf{Theme} & \textbf{Description} & \textbf{Key Codes} & \textbf{Count} \\
    \hline
    
    \multirow{1}{3cm}{Diverse SD Mitigation Practices}
        & \multirow{1}{4.5cm}{The range of tools and practices used to mitigate security vulnerabilities, showing how approaches vary across teams and what influences reliance on manual vs.\ automated solutions.}
        & Unaware of or no specialized tools & 7/22 \\
        \cline{3-4}
        & & Manual security reviews (e.g., log reviews, Jira tracking, manual vulnerability checks) & 3/22 \\
        \cline{3-4}
        & & Automated security tools (e.g., static analyzers, AWS CloudWatch/CloudTrail, custom scripts) & 5/22 \\
        \cline{3-4}
        & & Reliable vulnerability detection (e.g., internal/third-party detection tools, source code scanning, thorough testing) & 4/22 \\
        \hline
    
    \multirow{1}{3cm}{Cautious Adoption of AI in Security}
        & \multirow{1}{4.5cm}{How practitioners view the use of AI for SD mitigation, including emerging practices and concerns regarding trust, oversight, and limitations of current AI tools.}
        & Limited AI adoption & 13/22 \\
        \cline{3-4}
        & & Efficiency gains & 9/22 \\
        \cline{3-4}
        & & Human oversight required & 9/22 \\
        \cline{3-4}
        & & Privacy and trust concerns & 9/22 \\
        \hline
    
    \multirow{1}{3cm}{Gaps and Ongoing Challenges in SD Mitigation}
        & \multirow{1}{4.4cm}{Shortcomings in current mitigation practices, identifying where organizational workflows and responsibilities fall short during development.}
        & Lack of awareness of available tools or methods & 5/22 \\
        \cline{3-4}
        & & No active mitigation & 5/22 \\
        \cline{3-4}
        & & Reliance on security teams & 2/22 \\
        \cline{3-4}
        & & Code-level challenges & 2/22 \\
        \hline
    
    \end{tabular}
    }
    \label{RQ3_analysis}
\end{table}

\subsubsection{Diverse Vulnerability Mitigation Practices} 
\noindent This theme highlights the range of tools and practices practitioners use to mitigate security vulnerabilities. It shows how mitigation approaches vary across teams and projects, and what influences whether organizations rely on manual processes or automated solutions. \\

\noindent\textbf{Findings.}
While some participants did not use specific tools (P4, P6) or were unaware of available tools or methods for SD mitigation (P2, P12, P17, P18, P19), others reported using:
i) static methods (P1, P16), where, for instance, P1 stated the following: \quoteFont{``...static analysis tools...we develop our own''};
ii) manual checks such as reviewing logs and data visualization (P3), tracking SDs using Jira (P5), and checking for common code vulnerabilities like SQL injection and access control issues (P7):
\quoteFont{``Since I do not use automated security tools, I manually check for common vulnerabilities like SQL Injection and access control issues.''}; and
iii) automated tools. The latter include AWS CloudWatch and AWS CloudTrail for monitoring and tracking access logs (P9, P20),
as well as internal and third-party tools for detection (P11, P15). Other participants reported using static and dynamic analysis, custom security tools and scripts (P10), and code simulation practices (P20, P21). Some (P21, P22) highlighted thorough testing routines, source code scanning, and AI-based tools such as DeepCode to simulate and verify code security prior to deployment. \cb{This diversity of tools and practices indicates lack of standardized approaches to SD mitigation, with participants relying on a combination of manual checks and automated tools depending on context.}

\cor{Tool usage for SD mitigation varied widely across participants with different expertise and security experience. While some senior or security-focused participants (e.g., P15, P16) reported using advanced or layered tools, others in similar roles (e.g., P12, P17) were unaware of available methods. Conversely, several non-senior participants (e.g., P9, P20--P22) described using robust automated or AI-based tools, while others were also unaware of available methods (e.g., P4, P6). This suggests that tool adoption is not strictly tied to role or experience, but may depend on organizational context, awareness, or personal initiative.}

\cor{Overall, while some participants were unaware of available SD mitigation tools and/or methods in their organizations, possibly hindering effective security management regardless of their security expertise or years of experience, the remaining participants inconsistently used various manual and automated SD approaches.}

\begin{takeawaybox}
\minorrev{\textbf{Diverse Vulnerability Mitigation Practices.}}
SD mitigation practices vary widely and inconsistently across participants, ranging from no tools or awareness to manual reviews and automated detection, with tool adoption shaped more by organizational context than by role or experience.
\end{takeawaybox}

\subsubsection{Cautious Adoption of AI in Security}
\noindent This theme reflects how practitioners view the use of AI for SD mitigation. It helps identify emerging practices and concerns, particularly regarding trust, oversight, and the limitations of current AI tools in handling sensitive security tasks. \\

\noindent\textbf{Findings.}
Most participants (P1--P7, P10, P11, P13, P14, P16, P17) did not utilize AI-based tools for mitigating SDs. A few practitioners (P8, P9, P12, P15), however, mentioned experimenting with AI assistance and observing some efficiency benefits. P9 shared: \\

\begin{myquote}
    {\quoteFont{
    ``...we use an internal AI-based tool instead of ChatGPT. It helps generate Identity Access Management policies and refines code security''
    }}
\end{myquote}

A few additional participants (P18--P22) mentioned using AI tools, particularly proprietary or enterprise-specific LLMs (e.g., GitHub Copilot, DeepCode, PyCharm Copilot) for tasks such as secure authentication, bug detection, and code validation. Although several participants acknowledged that these tools are efficient and fast, none of the participants fully trust the AI findings. The majority emphasized that manual validation is essential either before or after using AI tools, especially in sensitive security contexts, as P8 stated, \quoteFont{``human expertise is necessary for validating security measures.''}.
\cb{Nine out of the \total participants raised privacy and trust concerns as key barriers to AI adoption, particularly in security contexts where sensitive code and data are involved, as reflected by P9's preference for an internal AI-based tool rather than publicly available solutions such as ChatGPT.}
In particular, P21 expressed that, while AI tools generally offer high reliability, they rarely achieve full coverage of all security issues, noting that their effectiveness remains limited and still requires further improvement.
% \cb{Based on their experience, they estimated the effectiveness to be around 60\%–70\%.}

\cor{The use of AI-based tools for SD mitigation varies among participants regardless of their seniority or security expertise. In particular, while some participants (e.g., P8, P9, P12, P15, P18--P22) actively use AI tools, many others (e.g., P1--P7, P10, P11, P13, P14, P16, P17) do not rely on AI for SD mitigation. However, all participants consistently agree that AI findings cannot be fully trusted and require manual validation and human expertise, as emphasized by P8 and P21. This highlights a common cautious awareness of AI’s current limitations across all experience and expertise levels.
} 

Overall, most participants do not rely on AI tools for mitigating SDs. While a few of the participants who use AI in mitigating SDs reported efficiency gains, they consistently emphasized the importance of manual oversight to ensure the accuracy and reliability of AI-generated results, \cor{regardless of their years of experience and seniority level.}

\begin{takeawaybox}
\minorrev{\textbf{Cautious Adoption of AI in Security.}}
Most participants do not use AI for SD mitigation; those who do report efficiency gains but consistently emphasize that AI findings require manual validation, reflecting cautious awareness of AI limitations across all experience levels.
\end{takeawaybox}

\subsubsection{Gaps and Ongoing Challenges in Vulnerability Mitigation} \noindent This theme addresses shortcomings in current mitigation practices and helps identify where organizational workflows and responsibilities fall short, preventing vulnerabilities from being addressed effectively during development. \\

\noindent\textbf{Findings.}
Several participants (P2, P4, P6, P12, P17) revealed significant gaps in their security practices, admitting they had no formal process for addressing SDs. Others (P5, P8) indicated that they do not handle such vulnerabilities directly, instead relying on issue trackers or specialized security teams as P8 quoted {\quoteFont{``Security vulnerabilities are mainly handled by a dedicated security team''}}. 
This suggests that, in many cases, unaddressed security vulnerabilities are not actively mitigated during the development phase of the SDLC.
Other participants reported similar issues, with the absence of formal mitigation processes often leading to a dependency on external teams or tools, leaving unresolved security vulnerabilities in the development stage of the SDLC. 

\cor{Regardless of security expertise or seniority, several participants (e.g., P2, P4, P6, P12, P17) revealed significant gaps in formal SD mitigation processes, often leading to reliance on specialized security teams (e.g., P5, P8) and resulting in security vulnerabilities being unaddressed during the development phase of the SDLC.}

\begin{takeawaybox}
\minorrev{\textbf{Gaps and Ongoing Challenges in Vulnerability Mitigation.}}
Several participants lack formal SD mitigation processes regardless of seniority, often leading to reliance on specialized security teams and leaving vulnerabilities unaddressed during the development phase.
\end{takeawaybox}

\subsubsection*{Cross Background, Experience, and Regional Pattern Analysis}
\noindent
Table~\ref{RQ3_analysis} maps RQ3 findings against participant background,
Analyzing RQ3 findings against participant background reveals notably different patterns from RQ1 and RQ2. Regarding \textit{diverse mitigation practices}, no clear experience- or region-based pattern emerged, while some senior and security-focused participants (P15, P16) reported using advanced layered tools, others in similar roles (P12, P17) were unaware of available methods, and several junior and mid-level participants (P9, P21, P22) described robust automated and AI-based tools. This suggests that tool adoption is shaped more by organizational context than by experience or region. Concerning \textit{cautious adoption of AI in security}, no background-related divergence was observed, with non-use of AI tools spanned all experience tiers and regions (P1--P7, P10, P11, P13, P14, P16, P17), while those who did use AI (P8, P9, P12, P15, P18--P22) were equally distributed across experience levels and regions. Crucially, the requirement for manual validation was universally agreed upon regardless of seniority, role, or region. For \textit{gaps and ongoing challenges}, similarly no strong experience or regional pattern emerged, with formal mitigation gaps reported across junior, mid-level, and senior participants alike (P2, P4, P6, P12, P17), suggesting that organizational structure rather than individual background drives these shortcomings.

%\begin{takeawaybox}
%Unlike RQ1 and RQ2, RQ3 reveals that SD mitigation practices, AI adoption, and mitigation gaps show no consistent pattern by experience level or region.
%Tool adoption and formal process gaps are distributed across all backgrounds, suggesting that organizational context and awareness are stronger drivers of mitigation behavior than seniority or location. The one universal finding across all backgrounds is cautious reliance on AI, with manual validation considered essential regardless of experience or expertise.
%\end{takeawaybox}

\vspace{3pt}
\begin{summarybox}[RQ3 Summary]
    \minorrev{SD mitigation practices were inconsistent across participants, with limited adoption of dedicated tools and cautious use of AI-assisted approaches. Findings suggest that organizational context and security awareness influence mitigation efforts more strongly than practitioner role, experience, or region.}

    %Many participants \cor{with all seniority levels and expertise areas} lack consistent or tool-supported practices for mitigating SDs. While some use manual methods or basic automation, AI adoption remains limited and cautious, with strong reliance on human oversight. Teams frequently delegate security tasks, revealing a need for more integrated, developer-level responsibility and tooling.
    %\cb{Notably, no consistent pattern by experience level, role, or region was observed across RQ3 themes, suggesting that organizational context and awareness are stronger drivers of mitigation behavior than individual background.}
    %\cb{Specifically, 7/22 practitioners were entirely unaware of available mitigation tools, only 3/22 used manual security reviews, and automated tools were used by 5/22. AI adoption was limited (13/22 did not use it), and those who did estimated its effectiveness at only 60-70\%, with human oversight considered essential across all participants.}
\end{summarybox}

\subsection{\textbf{How do software practitioners communicate the risks associated with SDs in software to teammates and decision makers? (RQ4)}}\label{RQ4}

\noindent As illustrated in the three themes of Table~\ref{RQ4_analysis}, we examine how practitioners i) \cb{identify and report SDs during the SDLC}, ii) mitigate their impact on the user experience, and evaluate cases where fixing such issues improved system stability. \cb{For each theme, the table shows the key codes and the number of participants who contributed to each.}

%% NEW TABLE %%
\begin{table}[H]
    \renewcommand{\arraystretch}{1.25}
    \centering
    \caption{\cb{Themes of SD communication and handling strategies (RQ4)}}
    \resizebox{\textwidth}{!}{
    \begin{tabular}{p{3cm} p{5.75cm} p{5.8cm} p{1.5cm}}
    \hline
    \rowcolor{gray!50}
    \textbf{Theme} & \textbf{Description} & \textbf{Key Codes} & \textbf{Count} \\
    \hline
    
    \multirow{1}{3cm}{Identifying SDs During the SDLC}
        & \multirow{1}{5.75cm}{How practitioners detect and track SDs during the testing phase, and what role manual checks and automated tools play in identifying unresolved security issues before release.}
        & Manual tracking (e.g., Jira tickets, log reviews, testing reviews) & 8/22 \\
        \cline{3-4}
        & & Automated detection (e.g., mutation testing, monitoring tools, internal security checks) & 8/22 \\
        \cline{3-4}
        & & QA team coordination (e.g., Scrum-based testing, QA escalation before release) & 3/22 \\
        \hline
    
    \multirow{1}{3cm}{Preventing SDs from Impacting Users}
        & \multirow{1}{5.75cm}{Whether and how practitioners attempt to prevent SDs from affecting end users, including design, testing, and communication strategies to reduce user-facing consequences.}
        & Active mitigation strategies (e.g., code reviews, error handling, input controls) & 7/22 \\
        & & Outside role responsibilities & 6/22 \\
        \cline{3-4}
        & & Unavoidable impact in some cases & 3/22 \\
        \cline{3-4}
        \hline
    
    \multirow{1}{3cm}{Addressing SDs to Improve System Stability}
        & \multirow{1}{5.75cm}{Real-world cases where fixing SDs improved system reliability and performance, highlighting the practical benefits and occasional unintended consequences of proactive SD remediation.}
        & Impactful fixes improving security and stability & 7/22 \\
        \cline{3-4}
        & & Unintended consequences of fixes (e.g., restricted access, cross-product instability) & 2/22 \\
        \cline{3-4}
        & & No significant examples reported & 13/22 \\
        \hline
    \end{tabular}
    }
\label{RQ4_analysis}
\end{table}

\subsubsection{Identifying SDs During Testing}  
\noindent This theme explores how practitioners \cb{detect, report, and track} SDs during the \cb{SDLC}. This is useful for understanding when and how SDs are discovered, and what role manual checks and automated tools play in identifying unresolved security issues before release.

\vspace{4pt}
\noindent\textbf{Findings.}
Most participants (P1--P7, P9--P14, P16, P19--P22) discussed methods \cb{used to track and report SDs during the SDLC.} Several (P1, P3, P16, P5, P6, P20, P22) rely on manual tracking, mostly through tools like Jira (or similar platforms such as HubSpot) (P1, P3, P16, P20), with P3 stating, {\quoteFont{``Follow tickets in Jira for in-queue security tasks''}}.
Others (P5, P6, P22) highlighted manual methods such as testing reviews, ticketing, and internal meetings: {\quoteFont{``...meeting with QA team... internal meetings also with developers... communicate progress to the senior management team''}} (P5), and {\quoteFont{``...reviewing logs manually''}} (P4).

On the other hand, several participants (P1, P2, P7, P9, P11, P13, P20, P22) mentioned automated processes to detect and track SDs, including mutation testing (P2), implementing security checks using internal tools (P7, P11), and {``\quoteFont{monitoring tools and logging mechanisms''}} (P9). 
In the use of QA team coordination processes, P21 noted the use of QA team coordination processes such as Scrum \cor{to coordinate testing efforts sprint-wise, with testing done alongside development tasks and following security protocols. Further, P19 and P22 also emphasized the role of QA teams in managing and escalating unresolved vulnerabilities before release.}

\cor{Both senior and security-focused participants (e.g., P7, P14--P18) as well as those with non-senior roles or without a security focus (e.g., P1, P2, P9, P11, P22) use a mix of manual and automated, external and internal tools SD tracking methods, showing no clear preference based on seniority or security expertise, suggesting diverse practices tailored to context rather than role.}

Overall, participants \cor{with all levels of seniority and expertise areas} track SDs inconsistently, using both manual (e.g., Jira, testing reviews, internal meetings) and automated (e.g., mutation testing, monitoring and logging tools) methods, with varying follow-up and accountability processes.

\begin{takeawaybox}
\minorrev{\textbf{Identifying SDs During Testing.}}
Practitioners track SDs inconsistently during testing, using a mix of manual tracking, automated detection, and QA coordination, with no clear preference based on seniority or security expertise.
\end{takeawaybox}

\subsubsection{Preventing SDs from Impacting Users}  
\noindent This theme investigates whether and how practitioners attempt to prevent SDs from affecting end users. It helps uncover design, \cb{mitigation,} and communication strategies that aim to reduce \cb{or prevent} user-facing consequences of unresolved vulnerabilities. \\

\noindent\textbf{Findings.}
While some participants (P8, P11, P14, P15, P19,  P20) believed that ensuring user experience is unaffected by SDs falls outside their role responsibilities, some others (P1, P4, P6) felt it was impossible to fully shield the user experience from the effects of SDs.

In contrast, several participants (P3, P5, P7, P9, P10) described \cb{practices} such as internal meetings and code reviews \cb{to identify and manage security issues early.  Participants also shared examples such as improving error handling and applying input controls.} P21 emphasized optimizing error handling to avoid exposing backend issues and locking sensitive data to prevent leaks. P22 highlighted strict input controls to limit unexpected user actions. P19 noted the presence of customer-facing teams who assess client satisfaction with proposed design changes.

Some participants (P11, P13, P15) observed that security teams often prioritize addressing security concerns over preserving the user experience.
For example, P7 indicated taking steps toward rebuilding the system to improve both security and stability:

\vspace{4pt}
\begin{myquote}
    {\quoteFont{
    “I found a major security flaw in a project and recommended rebuilding the system. This improved security and system stability significantly.”
    }}
\end{myquote}
\vspace{4pt}

\cor{Regardless of seniority or security expertise, participants showed varied views on the impact of SDs on user experience. Some (e.g., P8, P11, P14, P15, P19, P20) saw it as outside their scope, others (e.g., P1, P4, P6) felt it is not always possible to prevent SDs from affecting the user experience, while a few (e.g., P3, P5, P7, P9, P10, P21, P22) actively implemented mitigation strategies. This suggests a nuanced understanding of SD impact on the software user experience, shaped more by team roles and organizational context than by seniority or security expertise.}

Participants \cor{from all expertise areas and experience levels} expressed diverse views on \emph{\textbf{preventing SDs from impacting users}}, notably through affecting product features or user experience. Some considered it impossible due to the nature of SDs, while others felt that the priorities of security teams often compromise user experience. A few saw it as outside the scope of their roles, whereas others proposed manual and automated solutions to mitigate potential impacts.

\begin{takeawaybox}
\minorrev{\textbf{Preventing SDs from Impacting Users.}}
Views on preventing SD impact on users vary: some consider it outside their scope, others see it as unavoidable, while a few actively implement mitigation strategies, shaped more by team context than by individual expertise.
\end{takeawaybox}

\subsubsection{Addressing SDs to Improve System Stability}  
\noindent This theme presents real-world cases where fixing SDs improved system reliability and performance, highlights the practical benefits of proactive SD remediation, and emphasizes its value for maintaining stable, secure software systems.

\vspace{4pt}
\noindent\textbf{Findings.}
Most participants (P1, P5, P9, P14--P16, P20--P22) discussed cases where fixing unaddressed security vulnerabilities significantly improved system security. For instance, fixing a data integrity issue (P1) prevented crashes, as P1 stated:

\vspace{4pt}
\begin{myquote}
    {\quoteFont{
    ``...data integrity issue...risk on software stability in the sense that it failed then crashed.''
    }}
\end{myquote}
\vspace{4pt}

Others shared that restricting overprovisioned AWS access (P9) reduced security exposure, while resolving a \texttt{robots.txt} vulnerability that unintentionally exposed sensitive directories (P15) prevented potential system compromise. Additionally, fixing a communication issue between embedded devices (P16) improved system functionality. P21 described how addressing a wallet hack required months of remediation following a severe breakdown that left critical data exposed. P22 highlighted a case where the implementation of stricter security controls unintentionally limited access to system components, making some endpoints unreachable. Similarly, P20 noted that resolving a security issue in one product affected the stability of related products.

\cor{There is no clear pattern based on security expertise or seniority. Senior and/or security-focused participants (e.g., P14–P16, P20–P22) as well as non-senior and/or non-security participants (e.g., P1, P5, P9) reported impactful cases where fixing SDs improved security or stability. This suggests that awareness of SD consequences is broadly distributed across roles.}

Based on the experiences shared by \cor{all} participants, we conclude that \cor{they recognize} that addressing SDs significantly enhances system security and stability, improving overall system functionality.

\begin{takeawaybox}
\minorrev{\textbf{Addressing SDs to Improve System Stability.}}
Practitioners broadly recognize that fixing SDs improves system security and stability, though some cases reveal unintended consequences such as restricted access or cross-product instability.
\end{takeawaybox}

\subsubsection*{Cross Background, Experience, and Regional Pattern Analysis}
\noindent
Table~\ref{RQ4_analysis} maps RQ4 findings against participant background, 
Analyzing RQ4 findings against participant background reveals that, consistently with RQ3, experience level and region show limited influence across all three themes. Regarding \textit{identifying SDs during testing}, both senior and security-focused participants (P7, P14--P18) and junior and mid-level participants (P1, P2, P9, P11, P22) used a comparable mix of manual and automated tracking methods, with no preference tied to seniority, role, or region, suggesting that tool choice is driven by organizational context rather than individual background. Concerning \textit{preventing SDs from impacting users}, participants across all experience tiers and regions held varied views. Some participants (P8, P11, P14, P15, P19, P20) considered it outside their role scope, while others (P1, P4, P6) viewed full prevention as impossible, and a few (P3, P5, P7, P9, P10, P21, P22) actively implemented mitigation strategies, with no discernible pattern by seniority or region, further confirming that team context and organizational structure shape these views more than individual background. For \textit{addressing SDs to improve system stability}, impactful remediation cases were reported by both senior and security-focused participants (P14--P16, P20) and non-senior participants (P1, P5, P9), indicating that awareness of SD consequences is broadly distributed across all backgrounds and regions.

\vspace{3pt}
\begin{summarybox}[RQ4 Summary]
    \minorrev{Practitioners employ diverse approaches to manage SD throughout the software development lifecycle, combining technical and collaborative practices. While views differed on whether users can be fully shielded from SD, participants generally agreed that remediation improves software security, stability, and reliability.}
    
\end{summarybox}

\section{Discussion}\label{sec:discussion}
This section i) reflects on our study findings in light of existing literature by critically positioning our results with respect to prior research, ii) discusses the practical implications of our study, derived from our analyses and findings (see Section~\ref{sec:findings}), iii) examines various threats to validity (internal, external, and construct), outlining their mitigation strategies, iv) presents limitations of the approach we followed and the corresponding findings, and v) explores directions for future work.

\subsection{Reflecting on Prior Research}
Below, we position our study within the existing literature by comparing the findings of our four research questions RQ1--RQ4 (see Section~\ref{objAndRQs}) with prior work. Specifically, we identify \cor{whether} our results confirm, refine, or diverge from earlier research on SDs, drawing attention to underexplored aspects in the literature. \cor{In particular, to validate RQ1 findings alignment with the literature, we focus on the understanding of SDs (what are they? how and why they occur in software?). To evaluate the alignment of our RQ2 findings with existing work, we focus on practitioners' behavior towards SDs. We then examine the mitigation strategies and tools identified in both our findings and the literature, as well as how SDs are communicated in both environments, to evaluate the alignment between the literature and our results, addressing RQ3 and RQ4, respectively.}

% \vspace{4pt}
\noindent\paragraph{\textbf{Definition of SD.}}
SD has been widely described as a security-specific variant of technical debt~\cite{rindell2019managing}. Prior studies define it as postponed or incomplete security-related work aimed at gaining short-term flexibility~\cite{huopio2020quest}. It is often framed as the cost of deferring essential security tasks due to \cor{i)} limited resources, such as lack of security tools~\cite{edmundson2022sans,rindell2019managing,odera2023security,siavvas2022technical,daneva2018security,diaz2024can}, insufficient security expertise~\cite{edmundson2022sans,ali2025assessing,coetzer2024managing,odera2023security}, budget constraints~\cite{edmundson2022sans,tondel2022influencing,odera2023security}, and lack of security testing~\cite{edmundson2022sans,odera2023security}; or \cor{ii)} strategic shortcuts~\cite{coetzer2024managing,kruke2024defining}. This debt includes unaddressed \cor{security} vulnerabilities and is recognized as a subset of technical debt by various sources~\cite{maier2017towards,rindell2019managing,siavvas2022technical,odera2023security}.

\cb{Based on our RQ1 findings, we propose the following practitioner-informed \emph{\textbf{definition of SD}}, comprising three complementary facets that jointly characterize the SD concept as follows:}
\begin{definitionbox}[Practitioner-Informed Definition of SD]
\begin{enumerate}[leftmargin=*, nosep]
    \item SD reflects design or implementation choices that hinder or have the potential to hinder security goals.
    \item SD represents the gap between a system's current security measures and its ideal security state, which evolves over time and context.
    \item SD includes postponed security decisions where elevated risk has been knowingly or unknowingly accepted.
\end{enumerate}
\end{definitionbox}

This goes beyond the generic notion of SD as ``\textit{accumulated security vulnerabilities}'' by introducing intentionality, contextual variability, and forward-looking risk.

\vspace{4pt}
\noindent\textbf{SD and Security Vulnerabilities.}
Previous research links SDs directly with security vulnerabilities, treating \minorrev{the latter} as central indicators of SD~\cite{kruke2024defining}. \cor{Security} vulnerabilities are generally defined as exploitable weaknesses in software that compromise the CIA triad.

Our \cor{RQ1 findings} clarify this connection with more precision, based on practitioner input, as follows.
\begin{itemize}
    \item[-] SD arises when known security vulnerabilities are postponed or left unresolved \cor{ unintentionally}.
    \item[-] \cor{Security} vulnerabilities with no known solution are not classified as SD, since no actionable deferral occurs.
    \item[-] SD may include non-vulnerability issues, such as architectural or access control weaknesses, often omitted in past treatments.
\end{itemize}
This helps establish clearer boundaries for what qualifies as SD.

\vspace{4pt}
\noindent\paragraph{\textbf{Root Causes of SD.}}
Prior research identifies several recurring causes of SD \cor{(see Section~\ref{state}) such as} lack of security awareness~\cite{huopio2020quest,tondel2022influencing,zhao2024identifying}, poor team security culture~\cite{odera2023security}, time/budget pressures~\cite{coetzer2024managing,edmundson2022sans}, lack of expertise~\cite{edmundson2022sans,odera2023security}, and unresolved technical gaps~\cite{ali2025assessing}.

Our study (RQ1 and RQ2) both confirms and extends these insights. The study identifies:
\begin{itemize}
    \item Time constraints and deadlines as the dominant cause, reinforcing findings from~\cite{tondel2022influencing, maier2017towards, diaz2024can}.
    \item Poor or missing security requirements, frequently overlooked due to prioritization of functional features~\cite{ali2025assessing,edmundson2022sans,rindell2019managing,maier2017towards}.
    \item Lack of expertise and insufficient training, aligning with~\cite{coetzer2024managing, odera2023security,edmundson2022sans,ali2025assessing}.
    \item Intentional security trade-offs, such as knowingly ignoring minor vulnerabilities to deliver on time, have not been reported in most previous studies.
    \item Configuration issues and access control flaws, which are highly specific implementation-level problems that past studies have not emphasized.
    \item Interestingly, our interview participants did not mention the lack of tools, which is often cited in academic work~\cite{edmundson2022sans,rindell2019managing,odera2023security,siavvas2022technical,daneva2018security,diaz2024can}. \cor{Further, our participants shared that code complexity and embedded system constraints are also \emph{\textbf{root causes of SDs}}, which has not been mentioned in the literature,} suggesting a \cor{disconnection} between academic assumptions and industry reality.
\end{itemize}

\vspace{4pt}
\noindent\paragraph{\textbf{Practitioner Behavior Toward SD.}}
Behavioral responses to SD have not been a central focus in almost all previous studies, to the best of our knowledge, which mainly emphasized tooling or \emph{\textbf{root causes of SDs}}, as indicated earlier.

\cor{Our RQ2 findings} shed light on how practitioners address SD, as follows.
\begin{itemize}
    \item Practitioners use sprint planning and backlogs to prioritize unresolved security items.
    \item Manual and automated tracking methods, including customized internal practices.
    \item Efforts to foster team communication around security goals.
\end{itemize}

This adds a behavioral perspective to the predominantly technical literature on SDs.

\vspace{4pt}
\noindent\paragraph{\textbf{Mitigation Strategies and Tools of SD.}}
Prior research highlights various \cor{preventive} mitigation approaches such as \cor{threat modeling-based techniques~\cite{konev2022survey,scandariato2015descriptive,maier2017towards}}, penetration testing, and automated platforms like Security Flama for DevOps integration~\cite{ali2025assessing}. These discussions tend to remain at a high level.

Our \cor{RQ3 findings} contribute a grounded, practitioner-reported view of mitigation tools, as follows.
\begin{itemize}
    \item Manual techniques: code inspection, log reviews, and data visualization.
    \item Tracking systems: Jira-based issue management.
    \item Specific tools: AWS CloudTrail
    % , Burp Suite, Nmap, Nessus, Qualys.
    \item AI and simulation: tools like DeepCode and in-house security simulators.
\end{itemize}
The responses offer richer detail and diversity than the generalized tool categories typically reported.

\vspace{4pt}
\noindent\textbf{Communication of SD Risks.}
Effective communication of SD risk is widely acknowledged as vital in prior work. \cor{Specifically,} studies underscore the need for coordination between development, operations, and security teams~\cite{edmundson2022sans}, structured communication frameworks~\cite{ali2025assessing}, and the usability of automated security findings~\cite{voggenreiter2024automated}. Developers also rely on community platforms like Stack Overflow to discuss security issues~\cite{edbert2023exploring}.

\cor{Our RQ4 findings} fill a notable gap by empirically examining how practitioners communicate SD risks, as follows.
\begin{itemize}
    \item They reveal how practitioners frame security risks to teammates and decision-makers.
    \item They show how informal or undocumented communication shapes security prioritization.
    \item They emphasize communication as a key factor in whether SD is eventually addressed, as supported by prior evidence~\cite{tondel2022influencing}.
\end{itemize}
This focus on risk communication complements prior work that emphasized technical documentation or tooling, but not everyday communication practices.

\begin{table*}[t]
\centering
\caption{\cb{Comparison with Related Studies}}
\label{tab:comparison}
\begin{tabular}{p{1.25cm} p{3.5cm} p{3.5cm} p{3.5cm}}
\hline
\textbf{Study} & \textbf{Main Findings} & \textbf{Alignment} & \textbf{Extension} \\
\hline

{\cite{kruke2024defining}} & SDs linked to vulnerabilities; driven by awareness gaps and time pressure 
& Confirms that time pressure and limited expertise contribute to SD accumulation 
& Shows how SDs are prioritized and communicated during development \\

{\cite{tondel2022influencing}} & Security experts influence prioritization (single organization) 
& Confirms role-based prioritization 
& Extends beyond single-organization setting to multiple roles and contexts \\

{\cite{diaz2024can}} & SDs linked to security risks in open-source projects 
& Confirms that unresolved SDs manifest as concrete security vulnerabilities in practice 
& Adds real-world decision-making and trade-offs (e.g., postponing fixes to meet deadlines) \\

{\cite{kudriavtseva2024you}} & Gap between academic metrics and industry use 
& Confirms limited use of formal metrics 
& Shows reliance on informal practices (e.g., backlog, communication) \\

{\cite{tony2022conversational}} & Tools support vulnerability fixing (student context) 
& Confirms that tools support vulnerability detection and fixing 
& Shows inconsistent real-world adoption and hybrid use \\

{\cite{voggenreiter2024automated}} & DevOps tools support communication and mitigation 
& Confirms that communication is critical for coordinating SD mitigation 
& Highlights informal and undocumented communication \\

{\cite{edmundson2022sans, coetzer2024managing, odera2023security}} & Causes: awareness, time pressure, expertise 
& Confirms that awareness gaps, time pressure, and limited expertise are key drivers of SDs 
& Identifies intentional trade-offs in practice \\

\hline
\end{tabular}
\end{table*}
\vspace{4pt}
\noindent\textbf{Reflection Summary}.
\cb{Table \ref{tab:comparison} provides a comparison of our findings with prior studies, highlighting areas of agreement as well as key differences in how SDs are understood and managed in practice. While our results confirm several known causes of SDs, they also offer additional insight into how practitioners make decisions, handle trade-offs, and communicate about security issues.}

This study offers a broad, empirically grounded view of SD across four critical dimensions: \cor{practitioner} conceptualization \cor{of SDs (RQ1)}, practitioner behavior \cor{towards SDs} (RQ2), \cor{SD} mitigation tools (RQ3), and communication practices (RQ4). Compared to earlier works that often focus narrowly on tool adoption~\cite{ali2025assessing}, small-sample case studies~\cite{diaz2024can}, or metrics~\cite{kudriavtseva2024you}, this work:
\begin{itemize}
    \item[-] draws on interviews with \total practitioners across domains and \cor{geographic locations}.
    \item[-] provides detailed, practice-oriented definitions and causes of SD.
    \item[-] surfaces real-world tool use and mitigation practices, including AI-enabled tools.
    \item[-] explores the role of informal and formal communication in SD management.
\end{itemize}
The study reinforces the need for integrated, security-aware development processes across the SDLC, particularly with respect to managing trade-offs between delivery, resource constraints, and the CIA triad.

\subsection{Practical Implications}
Our study highlights key implications for researchers, practitioners, \cb{and education}, by revealing perceptions, priorities, management\cor{, and communication} of SD in various contexts. It enriches academic discussions and provides practical guidance for enhancing security practices. We discuss below these implications and compare them with existing work on SD and vulnerabilities.

\subsubsection{Implications for Research}

\noindent\textbf{\cor{Beyond the Surface: A Deeper Look into SD}.}
Prior research on SD has been limited in scope and context, leaving many practical aspects under-explored~\cite{rindell2019managing,coetzer2024managing}. Our study addresses this gap by offering empirical insights from a diverse set of practitioners. 
Drawing on findings from RQ1 and \cor{RQ4}, which capture how SDs are perceived, managed, and communicated across teams, we provide a broader foundation for future research rooted in real-world experiences.
\cb{Specifically, RQ1 reveals that role and experience shape how practitioners define, identify, and prioritize SDs, suggesting that future research should investigate how these differences affect SD accumulation across teams and organizations. RQ2 shows that security trade-offs are mostly intentional and driven by delivery demands, yet no established models exist to support such decisions, pointing to the need for decision-support frameworks for security prioritization under time pressure. RQ3 highlights that AI adoption for SD mitigation remains limited and cautious, with privacy concerns and inconsistent tool usage as key barriers, suggesting that future work should examine how AI tools can be safely and effectively integrated into SD mitigation workflows.}
% \cb{In particular, future research should investigate how role and experience shape SD perception and decision-making, and develop models to support transparent, risk-aware SD trade-offs in software teams.}
% The findings also reveal new directions that warrant closer examination. 
% For instance, RQ2 shows that practitioners are especially concerned with implementation-level issues, such as misconfigured access controls and legacy configurations, which are also addressed in prior work~\cite{coetzer2024managing}. Moreover, RQ1 highlights that security trade-offs are mostly intentional, driven by delivery demands. These insights point to underexplored but practically significant topics, suggesting that future work should not only examine SD across varied organizational contexts, but also develop models to better understand and support these trade-offs in security decision-making. 
 \\

\noindent\textbf{Tools Are Not Always the Problem.}
While the literature often identifies lack of tools as a contributor to SD~\cite{coetzer2024managing}, findings from RQ2 show that none of the participants cited tool unavailability as a cause. Instead, time pressure, \cor{lack of experts/seniors to supervise junior staff}, and awareness \cor{of SD} were more commonly blamed. Furthermore, RQ3 shows that even with tools in place, teams rely heavily on human judgment to interpret results and drive remediation. This challenges assumptions in prior work and suggests research should explore tool adoption, usability, and integration rather than focusing solely on tool availability. At the same time, while our study includes \total participants from diverse countries and roles, further empirical validation is needed. Future work could build on our RQ1–RQ3 findings using large-scale surveys or mixed-methods designs to test generalizability. For instance, are the prioritization strategies in RQ2 shared across industries? Do other teams communicate SD risks as inconsistently as those in RQ4? Our findings offer a grounded starting point for such validation, with clear direction on where assumptions about tooling and behavior need to be revisited.

\subsubsection{Implications for Practitioners}

\noindent\textbf{Embedding Security into Planning and Team Communication.}
Findings from RQ2 show that developers often postpone security fixes under deadline pressure, even when they recognize the risks. This aligns with earlier findings~\cite{tondel2022influencing, rindell2019managing, coetzer2024managing} that emphasize how security is often sacrificed under delivery pressure, reflecting a common tension between delivering features quickly and maintaining secure systems. However, as shown in RQ4, these trade-offs are rarely made explicit or systematically discussed within teams. Instead, SD risks tend to be inconsistently communicated or deprioritized altogether. To address this, practitioners should establish communication norms around SD using regular syncs, documentation, and visible tracking systems (e.g., dashboards or backlog entries). Embedding SD into sprint planning and backlog grooming can help ensure that vulnerabilities are not only detected (as seen in RQ3) but also actively discussed and addressed before becoming systemic risks. This is in line with prior work~\cite{daneva2018security}, which emphasizes the importance of making security concerns explicit through practices like security-specific stories, training, and collaborative planning. These approaches help normalize discussions around security early in the process, strengthening awareness and reducing reactive fixes later in development.

\subsubsection{\cb{Implications for Education}}
\cb{Our findings reveal significant gaps in practitioners' awareness, skills, and knowledge of SD management tools, suggesting implications for computer science and software engineering education. 
SD concept and the associated causes, and implications on software quality need to be taught at institutions as an essential component of software engineering courses, especially at the undergraduate level, where students start to learn the basics of SDLC process.
This ensures that students understand both the theoretical foundations and practical consequences of accumulating SDs.
Further, educators should incorporate hands-on training with SD identification and management tools, using real-world case studies to bridge the gap between theory and practice, and demonstrate how SDs accumulate and impact software projects over time. 
Additionally, team projects should include components requiring students to identify, prioritize, document, and remediate SDs, developing essential skills before they dive deep in developing software. By addressing these educational needs during students' formal education, similar to secure coding practices, we can establish stronger SD management practices in the next generation of software practitioners to avoid the accumulation of unresolved security vulnerabilities in software.
}

\vspace{3pt}
\noindent\textbf{Consistent and Informed Use of Security Tools.}
Findings from RQ3 show that teams vary widely in how they apply and respond to security tools. Some use tools sporadically or as a checklist item, while others run scans regularly but ignore the results due to alert fatigue or lack of clarity. Practitioners emphasized that having tools is not enough; their value depends on consistent use, proper configuration, and human interpretation. Security champions or trained developers often serve as intermediaries between automated outputs and actionable decisions.
These observations suggest that organizations should not only standardize the use of security tools across the SDLC, but also ensure that teams are equipped to interpret the results meaningfully. Without this, tools risk becoming background noise or misapplied. This aligns with previous research~\cite{ali2025assessing,daneva2018security}, which shows that secure coding practices require reinforcement through training, routine reviews, and contextual awareness. Making tool outputs visible and actionable, rather than isolated or ignored, can support more consistent and effective mitigation of SD.

\color{black}

\subsection{Lessons Learned And Recommendations}
\cor{\cb{As shown in Table~\ref{lessons},} we illustrate lessons we learned from our study in SDs, and share the corresponding recommendations, where applicable.}
\begin{table}[ht]
\centering
\caption{Lessons Learned and Recommendations on SDs}\label{lessons}
\begin{tabular}{p{6.2cm} p{6.2cm}}
\rowcolor{gray!50}
\hline
\textbf{Lessons Learned} & \textbf{Recommendations} \\
\hline
SDs often arise from technical oversights, limited resources, and prioritizing delivery over security. &
Allocate dedicated resources and time for security alongside business and delivery goals. \\ \hline

SDs compromise the CIA triad, leading to persistent risks and expensive remediation. &
Conduct early and regular security risk assessments to protect confidentiality, integrity, and availability. \\ \hline

Security professionals generally manage SDs more effectively than others, but awareness and tool usage vary by role and experience. &
Offer targeted security training and encourage broader tool adoption, especially among non-security staff. \\ \hline

Managing SDs requires both proactive and reactive efforts, typically prioritized by severity and impact. &
Adopt prioritization techniques that rank SDs by potential impact and likelihood. \\ \hline

Resolving SDs requires balancing deadlines, security needs, and access to expertise. &
Foster cross-team coordination and involve security experts in development workflows. \\ \hline

Many teams lack standardized or tool-supported practices to mitigate SDs. &
Invest in integrated and automated tools to support consistent SD tracking and mitigation. \\ \hline

AI adoption in SD management is minimal, and strong human oversight remains essential. &
%Experiment with AI-assisted tools cautiously, maintaining strong human supervision. 
Leverage AI tools carefully ensuring that decisions are monitored and validated by human intervention.\\ \hline

Security is often treated as someone else's responsibility, with minimal developer involvement. &
Ensure security is a shared responsibility by actively involving developers through accessible training and tools. \\ \hline
Different teams handle SD tracking and coordination in inconsistent ways; while some do it manually, others automate the process. & Adopt a unified approach to SD tracking by integrating automation tools. \\ \hline
Fixing SDs improves not only security but also system stability and reliability. & Prioritize SD remediation  as a strategic initiative to ensure long-term software quality. \\ 
\bottomrule
\end{tabular}
\end{table}

\subsection{Threats to Validity}\label{sec:threats}

\subsubsection{Internal Validity} %threats to validity
\cb{The first author coded all interview transcripts using reflexive thematic analysis. Because coding is inherently interpretive, there is a potential for subjective bias. 
To mitigate this, the second and third authors reviewed the codes, and any disagreements or further clarification need were resolved through consensus meetings.} 
\minorrev{Nevertheless, as coding was not independently re-performed by the second and third authors, the frequency of code occurrences remained primarily determined by the first author. This was partially mitigated through meetings, where all authors discussed and refined the coding decisions collectively.}

\subsubsection{External Validity}

While the insights of the proposed qualitative analysis technique on SDs offer valuable depth, they are not intended for statistical generalization or broad application, given the restricted sample of \total participants. To mitigate this, we selected software practitioners from different institutions, occupying various roles and located in ten countries. 
Further, the non-statistical nature of our sampling limits generalizability beyond the studied contexts, so caution is needed when applying our findings and insights in different contexts. To mitigate this, we ensured diversity in participants’ roles, domains, and genders following established qualitative sampling guidelines~\cite{baltes2022sampling}, enhancing the relevance of our insights.

\subsubsection{Construct Validity} 

\cb{In our study, no formal member checking of the produced codes and themes was conducted with participants. Although we consistently sought clarification during interviews to ensure accurate data capture, participants were unavailable or uncommitted to follow-up meetings, preventing us from returning findings for validation. This may affect the confirmability of our thematic analysis, as participants' intended meanings may not be fully reflected in the final codes and themes. To mitigate this, the produced codes and themes were systematically reviewed and validated across all three authors through consensus meetings, ensuring that interpretations were not solely dependent on a single researcher's perspective.}

\cb{Additionally,} our participants come from different countries, backgrounds (e.g., software testers, software developers, penetration testers), and expertise levels, with varying proficiency in English, which may lead to differences in how they interpret and respond to the interview questions. This can impact the generalizability and relevance of the study’s findings.
To mitigate this, we reworded and further clarified some concepts in the asked questions, while preserving the exact semantics to guarantee consistency in answers.

Further, due to confidentiality concerns, we do not share direct participant quotations, which are typically used to demonstrate that themes accurately reflect participants’ perspectives. To mitigate this, we shared the full interview guide and provided a detailed, step-by-step description of our thematic analysis process, illustrated with a running example. These measures enhance transparency and support replicability.

\subsection{Limitations}\label{sec:limitations}
Although we followed TA~\cite{clarke2017thematic} and adhered to empirical standards~\cite{ralph2021acm} by %applying a qualitative analysis and 
evaluating collected data and sharing interview codes, defined themes, and participant quotations, there is still a risk of oversimplifying complex data and potentially missing some nuanced or contradictory insights.
\minorrev{In addition, as the second and third authors reviewed the coding output conducted by the first author rather than independently re-coding all transcripts, the frequency of code occurrences across transcripts remained primarily determined by the first author, which may have influenced the results.}
Further, despite ensuring rigorous data collection, fully understanding SDs is challenging due to their complexity and the different security practices adopted across organizations.

Although we did not use formal methods to validate that our interview questions sufficiently and fully cover the research questions, the questions for each research question were carefully aligned with and motivated by the corresponding research objectives. Specifically, we explained the use of each question addressing Research Question RQ1 and guided the reader through how RQ1 aligns with these questions. The same process was applied for the remaining RQs (RQ2--RQ4).

\subsection{Future Directions}\label{sec:future directions}
Despite the methodological approach we followed while conducting our interviews, the results cannot be generalized. As part of future work, we plan to design and deploy a follow-up survey, informed by our interview findings, to gather broader insights and enhance generalizability across a larger population of software practitioners from both industrial and open-source contexts. Moreover, future research should explore how organizational structures and development models (e.g., DevSecOps, Agile) shape SD-related decisions, investigate how our themes apply in specific domains such as healthcare or finance, and develop practical guidelines or tools to support SD recognition and communication in industrial settings.
Additionally, future research could explore the different aspects of SDs we touched upon in more depth, particularly by conducting longitudinal studies across different organizational types, allowing researchers to observe how SD aspects in this study evolve over time. \cb{Similarly, future work can investigate relationships such as (i) whether poor software quality leads to vulnerabilities, and (ii) whether security patches and mitigation efforts increase system complexity, reduce maintainability, or impact performance.}

\section{Conclusion}\label{sec:conclusion}
SDs represent critical challenges in software systems, often resulting in long-term risks and high remediation costs. In this paper, we conducted a qualitative empirical study to investigate how SDs are perceived, managed, and communicated by software practitioners. We interviewed \total participants from diverse organizations and roles, addressing four research questions: i) assessing practitioners' knowledge and awareness of SDs, ii) examining their behavior toward SDs in development, iii) exploring commonly used tools and strategies for mitigation, and iv) understanding how associated risks are communicated within teams and to decision-makers.
Our findings highlight the need to balance technical, resource, and business constraints while preserving the Confidentiality, Integrity, and Availability (CIA) of software systems. Practitioners should consistently adopt appropriate mitigation tools and methods, supported by both organizational practices and specialized expertise. Addressing SDs early and systematically is essential to improving system security, stability, and functionality, reinforcing the importance of integrated and proactive security practices across development teams.

\bibliography{bibliography}

\end{document}